\documentclass[dvips,a4paper,12pt]{article}
\usepackage{graphicx}
\usepackage{amsmath}
\usepackage{bm}
\usepackage{arydshln}
\usepackage{authblk}
\usepackage{gensymb} 
\usepackage[numbers,sort&compress]{natbib}
\usepackage[bookmarks=true,colorlinks=true,linkcolor=black,citecolor=magenta,breaklinks]{hyperref}

\allowdisplaybreaks  
\newcommand\TT{\rule{0pt}{2.5ex}}        
\newcommand\BB{\rule[-1.5ex]{0pt}{0pt}}  
\newcommand{\be}{\begin{equation}}
\newcommand{\en}{\end{equation}}
\newcommand{\bea}{\begin{eqnarray}}
\newcommand{\ena}{\end{eqnarray}}
\newcommand{\lbl}[1]{\label{eq:#1}}
\newcommand{\lbltab}[1]{\label{tab:#1}}

\newcommand{\rf}[1]{(\ref{eq:#1})}
\newcommand{\Table}[1]{\ref{tab:#1}}
\newcommand{\fig}[1]{\ref{fig:#1}}

\newcommand{\braque}[1]{{\langle #1 \rangle}}
\newcommand{\outleft}{%
\mathrel{\setbox1=\hbox{$\scriptstyle{out}$}\setbox0=\hbox{$\langle$}\copy0\kern-0.3\wd0\lower1.1\ht0\copy1\kern-0.4\wd1}}

\newcommand{\bc}{\begin{center}}
\newcommand{\ec}{\end{center}}
\newcommand{\bt}{\begin{tabular}}
\newcommand{\et}{\end{tabular}}
\newcommand{\ba}{\begin{array}}
\newcommand{\ea}{\end{array}}
\newcommand{\gapprox}{%
\mathrel{%
\setbox0=\hbox{$>$}\raise0.6ex\copy0\kern-\wd0\lower0.65ex\hbox{$\sim$}}}
\newcommand{\lapprox}{%
\mathrel{%
\setbox0=\hbox{$<$}\raise0.6ex\copy0\kern-\wd0\lower0.65ex\hbox{$\sim$}}}
\renewcommand{\slash}[1]{%
\mathrel{\setbox0=\hbox{$/$}\copy0\kern-\wd0\hbox{$#1$}}}
\newcommand{\im}{{\rm Im\,}}
\newcommand{\re}{{\rm Re\,}}

\newcommand{\meta}{m_\eta}
\newcommand{\metad}{m_\eta^2}

\newcommand{\mpi}{m_\pi}
\newcommand{\mpid}{m_\pi^2}
\newcommand{\fpid}{F_\pi^2}
\newcommand{\mkd}{m_K^2}

\newcommand{\disc}{\hbox{disc}}

\newcommand{\Kbar}{\bar{K}}

\newcommand{\undemi}{{\textstyle\frac{1}{2}}}

\renewcommand{\undemi}{{1\over2}}
\newcommand{\mundemi}{{-1\over2}}
\newcommand{\om}{Omn\`es\ }
\newcommand{\ket}[1]{\vert#1\rangle}

\newcommand{\etapi}{{\eta\pi}}
\newcommand{\Kp}{{K^+}}
\newcommand{\Km}{{K^-}}
\newcommand{\Kz}{{K^0}}
\newcommand{\Kzb}{{\bar{K}^0}}
\newcommand{\piz}{{\pi^0}}
\newcommand{\pip}{{\pi^+}}
\newcommand{\pim}{{\pi^-}}
\newcommand{\fpiq}{F_\pi^4}
\newcommand{\fpis}{F_\pi^6}

\title{ Extended chiral Khuri-Treiman formalism for
 $\eta\to 3\pi$ and the role of the $a_0(980)$, $f_0(980)$ resonances}

\author[1]{M. Albaladejo}
\author[2]{B. Moussallam}
\affil[1]{\small Instituto de F\'{\i}sica Corpuscular (IFIC), Centro Mixto
  CSIC-Universidad de Valencia, Spain}
\affil[2]{\small Groupe de Physique Th\'eorique, IPN (UMR8608), Universit\'e
  Paris-Sud 11, Orsay, France} 

\begin{document}

\date{\today}

\maketitle

\begin{abstract}
Recent experiments on $\eta\to 3\pi$ decays have provided an extremely precise
knowledge of the amplitudes across the Dalitz region which
represent stringent constraints on theoretical descriptions.
We reconsider an approach in which the low-energy chiral expansion is assumed
to be optimally convergent in an unphysical region surrounding the Adler zero,
and the amplitude in the physical region is uniquely deduced by an
analyticity-based extrapolation using the Khuri-Treiman dispersive formalism.
We present an extension of the usual formalism which implements the leading
inelastic effects from the $K\Kbar$ channel in the final-state $\pi\pi$
interaction as well as in the initial-state $\eta\pi$ interaction. 
The constructed amplitude has an enlarged region of validity and accounts in a
realistic way for the influence of the two light scalar resonances $f_0(980)$
and $a_0(980)$ in the dispersive integrals.
It is shown that the effect of these resonances in the low energy region of
the $\eta \to 3\pi$ decay is not negligible, in particular for the $3\pi^0$
mode, and improves the description of the energy variation across the Dalitz
plot.   
Some remarks are made on the scale dependence and the value of the double
quark mass ratio $Q$.
\end{abstract}
\tableofcontents

\section{Introduction}
The physics of QCD in the soft regime is dominated by the phenomenon of
spontaneous symmetry breaking because of the presence of three light quarks in
the standard model. The low-energy dynamics can then be
described accurately through an expansion built from a chiral effective theory
(e.g.~\cite{Leutwyler:2015jga} for a recent review). This approach, which
applies in both the Euclidean and Minkowski space-times is, to some extent,
complementary to the purely numerical lattice simulation method. In the
effective theory, however, part of the information on the non-perturbative QCD
dynamics is contained as sets of values of the chiral coupling
constants. These are not known, a priori, except for their order of
magnitude~\cite{Georgi:1992dw} and must be determined as part of the probing
of the effective theory.

In QCD, isospin breaking phenomena are driven by $m_d-m_u$, the mass
difference of the two lightest quarks. For low-energy observables, an isospin
breaking ratio which conveniently absorbs some of the next-to-leading (NLO)
chiral coupling constants was introduced in ref.~\cite{gl85} 
\be\lbl{Q2def}
Q^{-2}= \frac{m_d^2- m_u^2}{m_s^2 -((m_u+m_d)/2)^2}\ .
\en
In general, isospin breaking  effects induced by electromagnetism are
comparable in size to those proportional to $m_d-m_u$ and their precise
evaluation is made difficult by the poor knowledge of the associated chiral
coupling constants~\cite{Urech:1994hd}. In this respect, the $\eta \to 3\pi$
amplitude plays a special role because these electromagnetic contributions
vanish in the $SU(2)$ chiral limit~\cite{Sutherland:1966zz} and are thus
expected to be suppressed.  This has been confirmed in the work of
refs.~\cite{Baur:1995gc,Ditsche:2008cq} who evaluated the contributions
proportional to $e^2m_u$, $e^2m_d$.

A number of recent high-statistics experiments have studied the
$3\pi^0$ decay mode of the
$\eta$~\cite{Tippens:2001fm,Prakhov:2008ff,Unverzagt:2008ny,Adolph:2008vn,Ambrosinod:2010mj}
as well as the charged mode
$\pi^+\pi^-\pi^0$~\cite{Ambrosino:2008ht,Adlarson:2014aks,Ablikim:2015cmz,Anastasi:2016qvh}.
An extremely precise knowledge of the energy variation of the
amplitudes squared across the Dalitz plot, which are traditionally
represented by a set of polynomial parameters, has now become
available. These accurate experimental results allow for stringent
tests of the theoretical description of the amplitude which must
obviously be passed before one attempts to determine $Q$.
The Dalitz plot parameters derived directly from the NLO chiral amplitude,
which was first computed in~\cite{Gasser:1984pr}, are in clear disagreement
with experiment. 
For instance, the prediction for the parameter $\alpha$, involved in the
neutral mode, has the wrong sign. The same problems, essentially, are found in
the resummed expansion approach discussed recently in
ref.~\cite{Kolesar:2016jwe}.
The computation of the amplitude at the
next-to-next-to leading (NNLO) chiral order was
performed~\cite{Bijnens:2007pr}. The comparison with experiment again fails if
one assumes a simple naive model for the $O(p^6)$ couplings $C_i$ (classified
in ref.~\cite{Bijnens:1999sh}) which are involved as six independent
combinations. The $\eta$ decay amplitude thus contains crucial information on
the true QCD values of these couplings, which are essentially not known at
present. 

One obvious deficiency of the chiral expansion when calculating
scattering or decay amplitudes in physical regions is the lack of exact
unitarity (as emphasised e.g. in ref.~\cite{Truong:1988zp}) which is restored
gradually when going to higher orders. In the case of $\eta\to 3\pi$, a large
amount of work was devoted to the problem of estimating these unitarity, or
final-state rescattering, higher order
corrections~\cite{Neveu:1970tn,Roiesnel:1980gd,Kambor:1995yc,Anisovich:1996tx,Borasoy:2005du,Schneider:2010hs,Kampf:2011wr}.

In the present paper we reconsider, more specifically, the approach followed
in refs.~\cite{Neveu:1970tn,Kambor:1995yc,Anisovich:1996tx}. The main
underlying assumption is that the chiral expansion of the $\eta\to
\pip\pim\piz$ decay amplitude should be optimally converging in an
\emph{unphysical} region of the Mandelstam plane in the neighbourhood of the
Adler zero. The amplitude in the physical region is then deduced from a
well-defined extrapolation procedure based on the analyticity properties of
amplitudes in QCD, which utilise the set of dispersive equations derived
initially by Khuri and Treiman~\cite{Khuri:1960zz} and perfected in the work
of refs.~\cite{Neveu:1970tn,Kambor:1995yc,Anisovich:1996tx}. These equations
implement crossing-symmetry and unitarity in a more complete way than more
naive loop-resummation approaches~\cite{Borasoy:2005du}.

In previous work, $\pi\pi$ rescattering was assumed to be elastic. This is 
essentially exact in the physical $\eta$ decay region. However, the
dispersive formalism involves integrals over an energy range extending
up to infinity. A property of the $\pi\pi$ scattering amplitude in the
isoscalar S-wave is the sharp onset of $K\Kbar$ inelasticity associated with
the $f_0(980)$ scalar
resonance~\cite{AlstonGarnjost:1971kv,Protopopescu:1973sh}. Because of isospin
violation, the $\eta\pi\to \pi\pi$ amplitude actually exhibits a double
resonance effect from both the $f_0(980)$ and the $a_0(980)$
scalars~\cite{Achasov:1979xc} near the $K\Kbar$ threshold. Our aim is to
propose a generalisation of the Khuri-Treiman formalism which takes into
account $K\Kbar$ inelasticity in the unitarity relations for both $\pi\pi$
scattering and $\eta\pi$ scattering (which may be viewed as an initial-state
interaction).
Some approximations will be made, which simplify considerably the practical
implementation,  such that crossing-symmetry is maintained at
the level of the $\eta\to 3\pi$ amplitudes but not in the amplitudes involving
the $K\Kbar$ channel.
In this multichannel formalism, the double resonance effect is taken into
account in the dispersive integrals and, furthermore, the construction of the
$\eta\to 3\pi$ amplitude becomes valid in an extended energy region which
includes not only the $\eta$ decay region but also a portion of the $\eta\pi
\to \pi\pi$ scattering region. This will allow us to study how the energy
dependence induced by the two 1 GeV scalar resonances propagate down to the
low energy region and quantitatively affects the Dalitz plot parameters. The
fact that these resonances could be influential at low energy was pointed out
previously in ref.~\cite{AbdelRehim:2002an}.

The plan of the paper is as follows. We first review in sec. 2 the derivation
of the one-channel equations in which the amplitudes satisfy elastic $\pi\pi$
unitarity in the $S$ and the $P$-waves. In sec. 3 we write the unitarity
relations including the $\eta\pi$ and the $K\Kbar$ channels. A closed system
of unitarity equations involves, besides $\eta\to 3\pi$, isospin violating
components of $\eta\pi\to K\Kbar$, $\pi\pi\to K\Kbar$ as well as $K\Kbar\to
K\Kbar$ amplitudes. A multichannel set of Khuri-Treiman integral equations is
defined such that the solution amplitudes satisfy these unitarity
relations. We then discuss in sec. 4 the matching between the chiral expansion
amplitudes and the dispersive ones. We adopt a simple approach which consists
in imposing that the difference between the chiral NLO and dispersive
amplitudes vanishes at order $p^4$. This provides four equations which, in the
single-channel case determine completely the dispersive amplitude provided one
had introduced exactly four polynomial parameters in the Khuri-Treiman
representation. This is generalised to the multichannel situation, in which
one introduces 16 polynomial parameters. Finally, in sec. 5, the results on
the Dalitz plot parameters are presented and some remarks are made on the
determination of the quark mass ratio $Q$.

\section{Khuri-Treiman  equations in the elastic approximation}
We remind below how the dispersion relations-based equations
derived by Khuri and Treiman~\cite{Khuri:1960zz} for $K\to 3\pi$ decay
can be generalised and applied to $\eta\to 3\pi$,
following~\cite{Neveu:1970tn,Kambor:1995yc,Anisovich:1996tx}.

\subsection{Single-variable amplitudes for $\eta\to 3\pi$ }
As initially demonstrated in the case of $\pi\pi \to \pi\pi$
(see~\cite{Stern:1993rg}),  amplitudes involving four
pseudo-Goldstone bosons satisfy, in a certain
kinematical range, an approximate representation in terms of functions
of a single variable which have simple analyticity properties
(see~\cite{Zdrahal:2008bd} for a review).  
In the case of $\eta\to 3\pi$, and neglecting quadratic isospin breaking,
there are three functions involved~\cite{Kambor:1995yc,Anisovich:1996tx}:
$M_0$, $M_1$, $M_2$. We will follow the notation of
ref.~\cite{Anisovich:1996tx} and write the $\eta\to \pi^+\pi^-\pi^0$ amplitude
as
\be\lbl{eta3pidecomp}
\ba{l}
{\cal T}_{\eta\to \pi^+ \pi^- \pi^0}(s,t,u)\equiv A(s,t,u)
=-\epsilon_L \Big[M_0(s) -{2\over3}M_2(s)+ (s-u) M_1(t) \\[0.2cm]
\phantom{{\cal T}_{\eta\pi^0\to \pi^+ \pi^-}(s,t,u)=-\epsilon_L }
+(s-t) M_1(u)+M_2(t)+M_2(u) \Big]\ ,
\ea
\en
with an overall factor\footnote{It is convenient to formally factor
  out $\epsilon_L$ but the amplitude  is
  actually of the form: ${\cal T}_{\eta\to 3\pi}=
\epsilon_L\,{\cal A}+\Delta{m}^2_K\,{\cal B}+e^2\,{\cal C}$ where $e$ is
the electric charge and $\Delta{m}^2_K$ is the physical $K^0-K^+$ mass
squared difference (see appendix~\ref{sec:appendixA}).}
$\epsilon_L$ which is proportional to the isospin breaking
double quark mass ratio $Q^2$ given in eq.~\rf{Q2def} 
\be
\epsilon_L = {Q^{-2}}\, 
{\mkd-\mpid\over3\sqrt3\fpid}{\mkd\over\mpid}\ .
\en
The Mandelstam variables are defined, as usual, as
\be
s=(p_\pip+p_\pim)^2,\quad
t=(p_\eta-p_\pip)^2,\quad
u=(p_\eta-p_\pim)^2\ 
\en
and satisfy
\be\lbl{s+t+u}
s+t+u= 3 s_0,\quad s_0 =\frac{1}{3}\metad +\mpid\ .
\en

\begin{figure}
\centering
\includegraphics[width=0.70\linewidth]{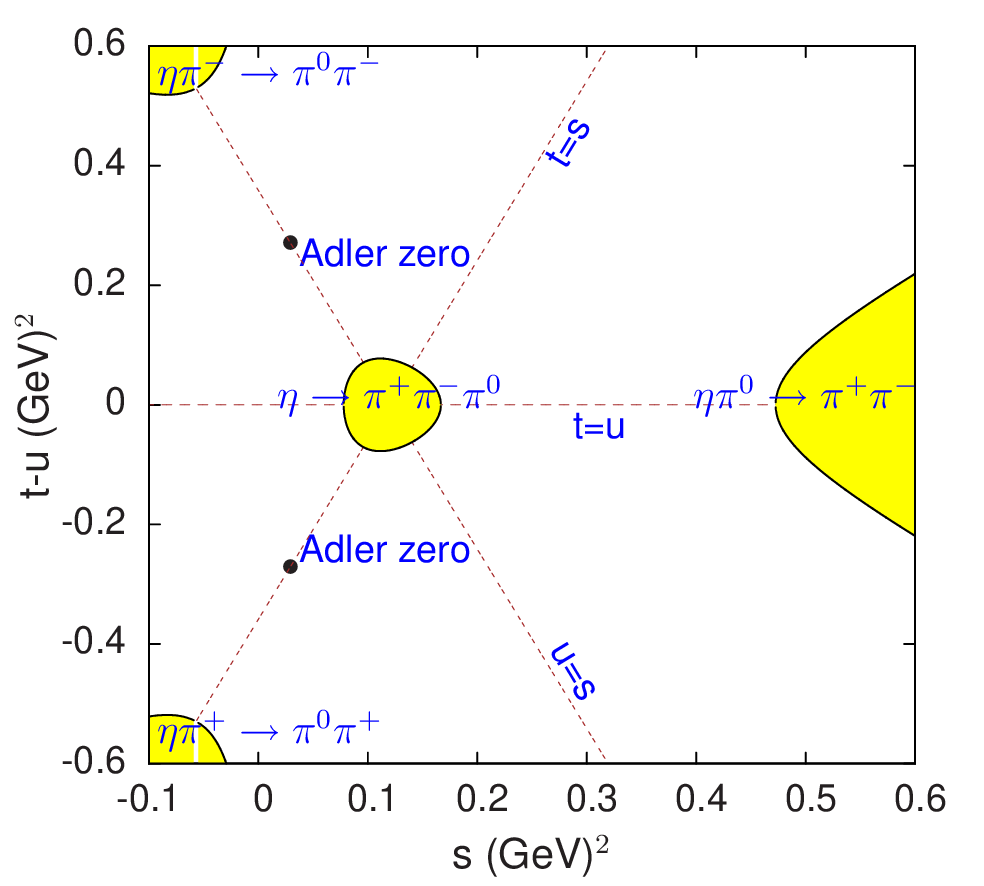}
\caption{\small Mandelstam plane showing the physical regions for $\eta \to
  3\pi$ decay and for $\eta\pi\to \pi\pi$ scattering. }
\label{fig:mandelstam}
\end{figure}

General analyticity properties imply that $\eta\to 3\pi$ decay and
$\eta\pi\to\pi\pi$ scattering are described by
the same function in different regions of the Mandelstam plane. The
corresponding physical regions are illustrated in fig.~\fig{mandelstam}. 
The analogous representation for $\eta\to 3\pi^0$ involves the two functions
$M_0$ and $M_2$ only and reads
\be\lbl{eta3pi0decomp}
{\cal T}_{\eta\to \pi^0\pi^0 \pi^0}(s,t,u)= -\epsilon_L \Big[
M_0(s) + M_0(t) +M_0(u) +\frac{4}{3}\big(
M_2(s) + M_2(t) +M_2(u)
\big)\Big]\ . 
\en

The representations~\rf{eta3pidecomp} and~\rf{eta3pi0decomp} are accurate in
regions of the Mandelstam plane where the imaginary parts of the partial-wave
amplitudes with angular momentum $j\ge 2$ in the $s$, $t$ or $u$ channels are
negligible (compared to those of the $j=0,1$ partial-waves).  In the case of
$\eta\to 3\pi$ or $\eta\pi \to \pi\pi$, this condition is satisfied in the
range where $s$, $t$, $u$ are sufficiently small compared with the masses
squared of the tensor resonances, i.e.  $|s|, |t|, |u|\lapprox 1$
$\hbox{GeV}^2$.  This condition is also satisfied exactly by the amplitude
obtained from the chiral expansion up to order $p^6$~\cite{Bijnens:2007pr}.
This will prove very useful for writing matching conditions.

The functions $M_I(w)$ are analytic in $w$ with a cut on the positive real
axis: $4\mpid \le w < \infty$. Based on Regge theory, we expect that the
functions $M_0(w)$ and $M_2(w)$ should not grow faster than $w$ at infinity,
while $M_1$ should be bounded by a constant. In the one-channel Khuri-Treiman
framework, both $M_0$ and $M_2$ are usually assumed to behave linearly in $w$
when $w\to\infty$.  With this asymptotic behaviour, there is a family of
redefinitions of the functions $M_1$ and $M_2$ which leaves the physical
amplitude $A(s,t,u)$ unmodified~\cite{Anisovich:1996tx,Lanz:2011},
\be\lbl{M1M2redef}
M_1(w) \to M_1(w) +a_1,\quad M_2(w) \to M_2(w) + a_2 + b_2\,w 
\en
($a_1$, $a_2$, $b_1$ being arbitrary constant parameters)
provided one correspondingly redefines $M_0$ as
\be\lbl{M0redef}
M_0(w) \to M_0(w) + a_0 + b_0\,w 
\en
with
\be
a_0=-\frac{4}{3}a_2+3s_0(a_1-b_2),\quad
b_0=-3a_1 +\frac{5}{3} b_2\ 
\en
and using the $s+t+u$ constraint~\rf{s+t+u}. 
This arbitrariness can be fixed by imposing the three $w=0$ conditions  
\be\lbl{MIzero}
M_1(0)=0\, ,\quad
M_2(0)=0\, ,\quad M'_2(0)=0\ .
\en
In the coupled-channel set-up, to be discussed below, the asymptotic condition
on $M_0$ will be modified such that $M_0(w)$ goes to a constant when $w\to
\infty$ instead of behaving linearly. This restricts the allowed
redefinitions of $M_1$, $M_2$  to those which satisfy $a_1=5/9\,b_2$. In that
case, only the two $w=0$ conditions $M_1(0)=M_2(0)=0$ can be imposed, while
$M'_2(0)$ is determined from the equations.  

These properties of the functions $M_I$
lead to the following dispersive representations
\bea\lbl{disprelMI}
&& M_0(s)=\tilde\alpha_0 + \tilde\beta_0 s
+{s^2\over\pi}\int_{4\mpid}^\infty ds'\,
\frac{\disc[M_0(s')]}
{(s')^2(s'-s)}         \nonumber\\
&& M_1(s)= {s\over\pi} \int_{4\mpid}^\infty ds'\,
\frac{\disc[M_1(s')]}
{s' (s'-s)}            \nonumber\\
&& M_2(s)= \tilde\beta_2 s+{s^2\over\pi}\int_{4\mpid}^\infty ds'\,
\frac{\disc[M_2(s')]}
{(s')^2(s'-s)} 
\ena 
defining
\be\lbl{discdef}
\disc[M_I(s)]\equiv {1\over 2i}( M_I(s+i\epsilon) -M_I(s-i\epsilon)
)\ .
\en

\subsection{Isospin amplitudes and crossing relations}
We choose the following conventional isospin assignment for the pions
and the kaons
\be\lbl{isoassign}
\begin{pmatrix}
\pi^+ \\    
\pi^0 \\    
\pi^- \\    
\end{pmatrix} \sim
\begin{pmatrix}
 -\ket{ 1\,1},\\                  
 \ket{ 1\,0},\\
 \ket{ 1-\!\!1}\\
\end{pmatrix}\ ,\quad               
\begin{pmatrix}
K^+\\[0.1cm]
K^0\\
\end{pmatrix}\sim
\begin{pmatrix}
\bar{K}^0\\[0.1cm]
-K^- \\
\end{pmatrix}\sim
\begin{pmatrix}
\ket{\undemi\,\undemi}\\[0.1cm]
\ket{\undemi\,\mundemi}\\
\end{pmatrix}\ .
\en
Let us consider the amplitudes which correspond to isospin states of
the $\pi\pi$ system,
\be
{\cal M}^{I,I_z}=\braque{\eta\pi \vert \hat{T} \vert \pi\pi;II_z}\ .
\en
One can express ${\cal T}_{\eta\pi^0\to \pi^+\pi^-}=A(s,t,u)$ in terms of $I=0,2$
isospin amplitudes, 
\be
A(s,t,u)=-{1\over\sqrt3}{\cal M}^{0,0}(s,t,u)
         -{1\over\sqrt6}{\cal M}^{2,0}(s,t,u)
\en
Further relations for the isospin amplitudes can be obtained using
crossing symmetries. Under $s-t$ and $s-u$ crossing one obtains,
\be\ba{l}
{\cal T}_{\eta\pi^- \to\pi^0\pi^-}=A(t,s,u)={1\over\sqrt2}(
{\cal  M}^{1,-1}(s,t,u)+ {\cal M}^{2,-1}(s,t,u))\\[0.2cm] 
{\cal T}_{\eta\pi^+\to\pi^+\pi^0}=A(u,t,s)={1\over\sqrt2}
({\cal M}^{1,1}(s,t,u)+ {\cal M}^{2,1}(s,t,u))\ .\\
\ea\en
Since the isospin breaking operator in QCD, $H_{IB}=
-1/2(m_d-m_u)\bar{\psi}\lambda_3\psi$,  transforms as
$I=1$,  $I_z=0$, one can use the  Wigner-Eckart theorem
\be\lbl{wignereckart}
\braque{j'm'\vert T^k_q \vert j m}=(-1)^{j'-m'}
\begin{pmatrix}
j' & k & j \\
-m' & q & m \\
\end{pmatrix}
\braque{j'\vert\vert T^k \vert\vert j} \ .
\en
which  yields the following relations among the ${\cal M}^{I,I_z}$ amplitudes
\be
\ba{l}
{\cal M}^{1,1}= -{\cal M}^{1,-1}\\
{\cal M}^{2,1}={\cal M}^{2,-1}={\sqrt3\over2} {\cal M}^{2,0}\ .
\ea
\en
One can then express the three independent isospin amplitudes  in
terms of the function $A(s,t,u)$
\be\ba{l}
{\cal M}^{0,0}(s,t,u)=  -\sqrt3\left( 
A(s,t,u)+{1\over3}( A(t,s,u)+A(u,t,s)) \right)\\[0.2cm]
{\cal M}^{1,1}(s,t,u)={1\over\sqrt2}\left( 
-A(t,s,u)+A(u,t,s)\right)\\[0.2cm]
{\cal M}^{2,1}(s,t,u)={1\over\sqrt2}\left( 
A(t,s,u)+A(u,t,s)\right)\ .\\
\ea\en
In the following we will simply denote
\be
{\cal M}^{0,0}\equiv {\cal M}^{0},\quad
{\cal M}^{1,1}\equiv {\cal M}^{1},\quad
{\cal M}^{2,1}\equiv {\cal M}^{2}\ .
\en
Inserting the  representation~\rf{eta3pidecomp} we
obtain an expression of the three isospin amplitudes in terms of the
one-variable functions $M_I(w)$
\be
\ba{l@{}r@{}l}\lbl{calMIdecomp}
{\cal M}^0(s,t,u)=& \sqrt3\,\epsilon_L & \left[ M_0(s)+ 
  {1\over3}M_0(t)+{10\over9}M_2(t)+{2\over3}(s-u)M_1(t)
+(t\leftrightarrow u) \right] \\[0.2cm]
{\cal M}^1(s,t,u)=&  {\sqrt2\over3}\,\epsilon_L & \left[ 
3t M_1(s)+{3\over2} M_0(t)+{3\over2}(s-u)M_1(t)-{5\over2}M_2(t)
-(t\leftrightarrow u\right] \\[0.2cm]
{\cal M}^2(s,t,u)=& -\sqrt2\,\epsilon_L & \left[ M_2(s)+ 
  {1\over2}M_0(t)+{1\over6}M_2(t)-{1\over2}(s-u)M_1(t) 
+(t\leftrightarrow u) \right] \\[0.2cm]
\ea\en

\subsection{Partial-waves and elastic $\pi\pi$ unitarity relations}
In order to derive expressions for the discontinuities
$\disc[M_I(s)]$, we must consider partial-waves and their unitarity
relations. We can define the partial-wave expansion of the isospin amplitudes
as 
\be
{\cal M}^I(s,t,u)=16\pi\sqrt2 \sum_j (2j+1) {\cal M}^I_j(s) P_j(z)
\en
where $z$ is the cosine of the scattering angle in the centre-of-mass
frame of $\eta\pi\to \pi\pi$ which is related to the Mandelstam
variables by
\be\lbl{tuandkappa}
t, u=\frac{1}{2}\left( \metad+3\mpid -s \pm \kappa(s)\,z\right),\quad
\kappa(s)=\sqrt{(1-4m^2_\pi/s)\,\lambda_{\eta\pi}(s)}
\en
with
\be
\lambda_{PQ}(s)=(s-(\meta+\mpi)^2)(s-(\meta-\mpi)^2)\ .
\en
From the representation of the isospin amplitudes,
eq.~\rf{calMIdecomp} one easily derives the expression for the partial-waves.
The result, for the $j=0,1$ partial-waves, can be written as, 
\be\lbl{calM00LR}
\ba{l}
{\cal M}^0_0(s)=\dfrac{\sqrt6\,\epsilon_L }{32\pi}
\,[ M_0(s) +\hat{M}_0(s) ]\\[0.3cm]
{\cal M}^1_1(s)=\dfrac{\epsilon_L }{48\pi}\,\kappa(s)\,[  M_1(s)+
  \hat{M}_1(s)]\\[0.3cm] 
{\cal M}^2_0(s) =-\dfrac{\epsilon_L }{16\pi}
\,[ M_2(s) +\hat{M}_2(s) ]\\
\ea\en
where the functions $\hat{M}_I(s)$ are given by linear combinations of
angular integrals of the functions $M_I$
\be\lbl{Mhatexpr}
\ba{rl}
\hat{M}_0=&{2\over3} \braque{M_0}+{20\over9}  \braque{M_2} +2(s-s_0)\braque{M_1}
+{2\over3}\kappa \braque{z M_1}\\[0.2cm]
\kappa\hat{M}_1=&3\braque{zM_0}-5\braque{zM_2}+{9\over2}(s-s_0)\braque{zM_1}
+{3\over2}\kappa\braque{z^2 M_1}\\[0.2cm]
\hat{M}_2=&\braque{M_0}+{1\over3}\braque{M_2}-{3\over2}(s-s_0)\braque{M_1}
-{1\over2}\kappa\braque{z M_1}\ .\\
\ea\en
with the notation~\cite{Anisovich:1996tx}
\be\lbl{angulint}
\braque{z^n M_I}(s)={1\over 2}\int_{-1}^1 dz\,z^n M_I(t(s,z) )\ .
\en

Writing unitarity relations, one must first consider the unphysical situation
where the $\eta$ meson is stable, $\meta \le 3\mpi$. The physical case is
defined using analytic continuation in $\meta$ as in the 
classic derivation of generalised unitarity~\cite{Mandelstam:1960zz}.
The contribution from the $\pi\pi$ states to the unitarity relation for the
partial-wave ${\cal M}_j^I$, reads
\be\lbl{unitrelelast}
\im {\cal M}_j^I(s)= \disc[ {\cal M}_j^I(s) ]=
\sigma_\pi(s) (f_j^I(s))^* {\cal M}_j^I(s)\ .
\en
with
\be
\sigma_P(s)=\sqrt{1-\frac{4 m_P^2}{s}}\,\theta(s-4m_P^2)\ 
\en
and $f_j^I(s)$ is the $\pi\pi\to\pi\pi$ partial-wave amplitude, which is
related to the scattering phase-shift by
\be
\exp(2i\delta_j^I(s))=1+2i\sigma_\pi(s) f_j^I(s)\ .
\en 
The equality between the imaginary part and the discontinuity in
eq.~\rf{unitrelelast} holds when $m_\eta < 3m_\pi$. For the physical value of
$m_\eta$ the right-hand side of eq.~\rf{unitrelelast} continues to give the
discontinuity across the unitarity cut, while the imaginary part must be
deduced from the dispersive representation. 

\subsection{Khuri-Treiman equations in the elastic approximation}
In the unphysical situation when $\meta < 3\mpi$, the cuts of the functions
$\hat{M}_I(w)$ are located in the region $Re[w]< 4\mpid$ such that
eqs.~\rf{calM00LR} correspond to a splitting of the partial wave amplitudes
into two functions which have a separated cut structure.  When the $\eta$ mass
is increased to its physical value, the $m_\eta^2+i\epsilon$ prescription must
be used~\cite{Mandelstam:1960zz} and this insures that the complex cut of the
$\hat{M}_I$ functions, which approaches infinitesimally close to the unitarity
cut in the region $4\mpid \le s \le (\meta-\mpi)^2$, remains well separated
from it (see fig.~4 in ref.~\cite{Descotes-Genon:2014tla}).  Using the fact
that $\hat{M}_I(s)$ has no discontinuity across the unitarity cut one can
deduce from~\rf{unitrelelast} that the discontinuities of the functions
$M_I(s)$ along $4\mpid \le s < \infty$ are given by
\be\lbl{discMI}
\disc[ M_I(s)]=  \exp(-i\delta_j^I(s))\sin\delta_j^I(s) [ M_I(s+i\epsilon) +
  \hat{M}_I(s)]
\en
where $j=0$ when $I=0,2$ and $j=1$ when $I=1$. In the sequel, we will
drop the $j$ subscript in the $\pi\pi$ phase-shift.
Eqs.~\rf{discMI} imply that the functions $M_I$ can be written as
Muskhelishvili-Omn\`es (MO) integral representations
\bea\lbl{dispomnes}
&& {M_0(s)}=  \Omega_0(s)\Big[ \alpha_0 +\beta_0 s 
+(\gamma_0 +\hat{I}_0(s))\,s^2 
 \Big]\nonumber\\
&& {M_1(s) }= \Omega_1(s)\Big[  (\beta_1 + \hat{I}_1(s))\,s 
 \Big] \nonumber\\
&& {M_2(s) }= \Omega_2(s)\Big[ \hat{I}_2(s)\, s^2
\Big]\ , 
\ena
where
\be
\hat{I}_a(s)=\frac{1}{\pi}\int_{4\mpid}^\infty ds'\,
\frac{\sin\delta^a(s') \hat{M}_a(s')}
{\vert\Omega_a(s') \vert (s')^{2-n_a}(s'-s)},\
n_a=\delta_{1a} 
\en
and where the \om functions $\Omega_I$ are given in terms of the
$\pi\pi$ phase-shifts by the usual relation
\be\lbl{Omegdef}
\Omega_I(s)=\exp\Big[
\dfrac{s}{\pi}\int_{4\mpid}^\infty ds' {\delta^I(s')\over (s')(s'-s)}
\Big]\ .
\en
The phase-shifts, in the present context, are usually taken to obey the
following asymptotic conditions 
\be\lbl{asyphaseshifts}
\delta^0(\infty)=\delta^1(\infty)=\pi,\quad 
\delta^2(\infty)=0
\en
which seem rather natural since these conditions are roughly satisfied at the
$K\Kbar$ threshold. In the elastic approximation framework, one can thus take
the phases to be constant or quasi-constant in the inelastic region, above 1
GeV. 

The polynomial part in the MO representation~\rf{dispomnes} was chosen to have
four parameters. This allows one to define a unique dispersive amplitude by
implementing four independent matching conditions with the chiral NLO
amplitude. Taking into account the asymptotic conditions on the
phase-shifts~\rf{asyphaseshifts} one easily sees that the Khuri-Treiman
equations~\rf{dispomnes} implement the following asymptotic behaviour for the
functions $M_I$ 
\be
M_0(w)\sim M_2(w)\sim w,\quad
M_1(w)\sim constant\ .
\en
The functions  $M_I$ also satisfy the $w=0$ conditions~\rf{MIzero}
and are thus uniquely defined.

\section{Beyond the elastic $\pi\pi$ approximation}
Elastic unitarity for $\pi\pi$ scattering is valid exactly below the four
pions threshold and approximately up to the $K\Kbar$ threshold. Close to 1
GeV, the $\pi\pi$ phase-shift increases very sharply under the influence of
the $f_0(980)$ resonance which also couples strongly to $K\Kbar$. In order to
properly account for the effect of this resonance it is thus necessary to go
beyond the elastic unitarity approximation. We  discuss in this section how to
include both the $\eta\pi$ and the $K\bar{K}$ channels into the unitarity
relations and then generate a generalisation of the Khuri-Treiman
equations. This will allow us to account for both the $f_0(980)$ and the
$a_0(980)$ resonances in a realistic way.

\subsection{ $\eta\pi$ contribution to unitarity}
Including the $\eta\pi$ channel in addition to $\pi\pi$, the unitarity
relation becomes
\be\lbl{unitrel+pieta}
\disc[ {\cal M}_j^I(s) ]=
 \sigma_\pi(s)     (f_j^I(s))^*        {\cal M}_j^I(s)
+\sigma_{\etapi}(s) ({\cal M}_j^I(s))^* {f}_j^\etapi(s)
\en
where
\be
\sigma_{PQ}(s)=\theta(s-(m_P+m_Q)^2)
      {\sqrt{\lambda_{PQ}(s)}\over s}
\en
and ${f}_j^\etapi(s)$ is the $\eta\pi\to \eta\pi$ partial-wave amplitude. In
the energy region where $\eta\pi$ scattering is elastic, which we will assume
to extend up to the $K\Kbar$ threshold, ${f}_j^\etapi(s)$ is related to the
scattering phase-shift by
\be
\exp(2i\delta_j^{\eta\pi}(s))= 1+2i \sigma_{\eta\pi}(s) {f}_j^\etapi(s)\ .
\en
The $j=1$ partial-wave ${f}_1^\etapi$ corresponds to exotic quantum numbers
$j^{PC}=1^{-+}$ and should thus remain rather small up to the 1 GeV
region\footnote{A resonance possibly exists in this
  amplitude~\cite{Adams:2006sa} with a mass $M\simeq 1.3$ GeV}.  Therefore,
$\eta\pi$ rescattering is expected to affect mainly the two $j=0$ amplitudes
${\cal M}_0^0$ and ${\cal M}_0^2$. In the elastic regime, using
eq.~\rf{unitrel+pieta} one easily derives that the relation between the
amplitudes on both sides of the unitarity cut reads,
\be
{\cal M}_0^I(s-i\epsilon) =
\exp(-2i\delta^I(s))
\exp(-2i\delta_0^{\eta\pi}(s)) {\cal M}_0^I(s+i\epsilon)\ ,\quad
I=0,2\ .
\en
Comparing with the analogous relation in the elastic unitarity
case~\rf{unitrelelast} one deduces that including the effect of $\eta\pi$
rescattering in the $\eta\pi\to\pi\pi$ amplitude (which can be viewed as an
initial-state interaction) amounts to simply perform the
following replacements in the \om representations~\rf{dispomnes},
\be
\delta^I(s) \to \delta^I(s)+\delta^{\eta\pi}_0(s)\ , \quad I=0,2\ .
\en
In practice, $\eta\pi$ rescattering is expected to become significant when the
energy approaches the mass of the $a_0(980)$ resonance. It becomes necessary,
then, to also take into account the $K\Kbar$ channel.

\subsection{ $K\Kbar$ contributions to unitarity}
Let us now include the $K\Kbar$ states into the partial-wave unitarity
relations.  
\begin{itemize}
\item[a)]$I=1, j=1$:
We are concerned mainly with $j=0$ amplitudes but let us consider
the $j=1$ amplitude ${\cal M}_1^1$ here also for completeness. The $K\Kbar$
contribution reads
\be
\disc[ {\cal M}_1^1(s) ]_{KK} = 
\sigma_{K^+K^0}(s) 
\left( T_1^{\Kp\Kzb \to \pi^+\pi^0 }(s)\right)^* 
       T_1^{\eta\pi^+ \to \Kp \Kzb}  (s)         \ .
\en
The amplitude $T_1^{\eta\pi^+ \to \Kp \Kzb}$ is isospin
violating\footnote{ The amplitudes $T_j^{\eta\pi^+\to \Kp\Kzb}$ are isospin
 violating (conserving) for odd (even) values of $j$.  This can be seen using
 G-parity: $G\ket{(K\Kbar)^I_j }= (-1)^{I+j} \ket{(K\Kbar)^I_j }$.  Since
 $I=1$ for $K^+\Kzb$, $G=+1$ for odd values of $j$ and $-1$ for even values
 while $G=-1$ for $\eta\pi$.}  and, at linear order in isospin breaking, one
can set $\sigma_{K^+K^0}(s)=\sigma_K(s)$.  We will denote the amplitudes
appearing above as
\be
T_1^{\Kp\Kzb \to \pi^+\pi^0 }(s) \equiv g^1_1(s)\ ,\quad
T_1^{\eta\pi^+ \to \Kp \Kzb}(s)\equiv {\cal N}^1_1(s) \ .
\en
\item[b)]$I=2, j=0$:
For the ${\cal M}_0^2$  amplitude now, the $K\Kbar$ contribution reads
\be\lbl{discI=2}
\disc[{\cal M}_0^2(s) ]_{KK} = 
\sigma_{K^+K^0}(s)  
\left( T_0^{\Kp\Kzb \to \pi^+\pi^0 }(s)\right)^* 
       T_0^{\eta\pi^+ \to \Kp \Kzb}(s)            
\en
The amplitude $T_0^{\eta\pi^+ \to \Kp \Kz} $ is isospin conserving in 
this case, since $j$ is even while the amplitude 
$T_0^{\Kp\Kzb \to \pi^+\pi^0} $ is isospin violating. 
We will denote the two amplitudes in eq.~\rf{discI=2} as
\be\lbl{denoteI=2}
T_0^{\eta\pi^+ \to \Kp \Kzb}(s) \equiv g_0^\etapi(s)\ ,\quad
T_0^{\Kp\Kzb \to \pi^+\pi^0}(s) \equiv  {\cal G}_0^{12}(s)\ .
\en
\item[c)]$I=0, j=0$:
Finally, for the ${\cal M}_0^0$  amplitude, the $K\Kbar$ contributions to the
unitarity relations are
\be
\ba{ll}
\disc[ {\cal M}_0^0(s) ]_{KK} &=
\sigma_\Kp(s) 
\left(T_0^{\Kp\Km\to (\pi\pi)^0}(s)  \right)^*  
      T_0^{\eta\pi^0\to\Kp\Km}(s)          \\[0.2cm]
\ &+\sigma_\Kz(s) 
\left(T_0^{\Kz\Kzb\to (\pi\pi)^0}(s)  \right)^*  
      T_0^{\eta\pi^0\to\Kz\Kzb}(s)          \\
\ea\en
\end{itemize}
Let us now separate the isospin conserving and  the isospin violating
contributions. For the kinematical factors, we introduce
\be\lbl{Deltasig}
\sigma_K(s)=\undemi( \sigma_\Kp(s) +\sigma_\Kz(s))\ ,\quad
\Delta\sigma_K=\undemi( \sigma_\Kp(s) -\sigma_\Kz(s))\
\en
For the $\eta\piz \to \Kp\Km,\, \Kz\Kzb$ amplitudes, the isospin conserving
part, $g^\etapi_0$, was introduced in eq.~\rf{denoteI=2}
and we call the isospin violating one ${\cal N}^0_0$. One has 
\be\ba{l}
g^\etapi_0(s) = {1\over\sqrt2}\left( T_0^{\eta\piz\to\Kp\Km}(s)-
T_0^{\eta\piz\to\Kz\Kzb}(s)\right)\\[0.2cm] 
{\cal N}^0_0(s)= {1\over\sqrt2}\left( T_0^{\eta\piz\to\Kp\Km}(s)+
T_0^{\eta\piz\to\Kz\Kzb}(s)\right)
\ea\en
For the $K\Kbar\to (\pi\pi)^0$ amplitudes, the  isospin conserving
and the isospin violating amplitudes are denoted as
\be\ba{ll}
I=0:\quad & g_0^0  = {1\over\sqrt2}\left( T_0^{\Kp\Km\to (\pi\pi)^0} +
T_0^{\Kz\Kzb\to (\pi\pi)^0}\right) \\[0.2cm]
I=1\to I=0:\ & {\cal G}^{10}_0 ={1\over\sqrt2}\left( T_0^{\Kp\Km\to (\pi\pi)^0} -  
T_0^{\Kz\Kzb\to (\pi\pi)^0}\right) \\
\ea\en
Using this notation, the $K\Kbar$ contributions in the unitarity relations for
the partial-waves ${\cal M}^I_j$ can be summarised as follows
\be\ba{lll}
I=0:\ & \disc[ {\cal M}_0^0(s) ]_{KK} &=\sigma_K(s) 
\left[ \left({\cal G}^{10}_0(s)\right)^*  g^\etapi_0(s) 
+      \left(            g^0_0(s)\right)^* {\cal N}^0_0(s) \right] \\[0.2cm]
\ &\ &+\Delta\sigma_K(s)
\left[ \left(  g^0_0(s) \right)^*\,g^\etapi_0(s)\right]            \\[0.2cm]
I=1:\ & \disc[ {\cal M}_1^1(s) ]_{KK} &=\sigma_K(s)
\left(g_1^1(s)\right)^*\,{\cal N}_1^1(s)                         \\[0.2cm]
I=2:\ & \disc[ {\cal M}_0^2(s) ]_{KK} &=\sigma_K(s) 
\left({\cal G}^{12}_0(s)\right)^*   g^\etapi_0(s) \ .\\
\ea\en
These contributions involve new isospin-breaking $K\Kbar\to \pi\pi$ and
$\eta\pi\to K\Kbar$ amplitudes: ${\cal G}^{10}_0$, ${\cal G}^{12}_0$, ${\cal
  N}^0_0$, ${\cal N}^1_1$.  In order to write a closed set of unitarity equations
we must also consider $K\Kbar\to K\Kbar$ amplitudes. For these, the isospin
conserving/violating components are denoted  $h_j^I$, ${\cal H}^{10}_j$
\be\ba{ll}
T_j^{\Kp\Km\to\Kp\Km}(s) &= 
{1\over2}\left( h_j^0(s)+h_j^1(s)+2{\cal H}^{10}_j(s)\right)\\[0.2cm]
T_j^{\Kp\Km\to\Kz\Kzb}(s) &= 
{1\over2}\left( h_j^0(s)-h_j^1(s)\right) \\[0.2cm]
T_j^{\Kz\Kzb\to\Kz\Kzb}(s) &=
{1\over2}\left( h_j^0(s)+h_j^1(s)-2{\cal H}^{10}_j(s)\right)\ .\\
\ea\en
Table~\Table{notations} summarises our notation for the various amplitudes
involved. 
\begin{table}[hbt]
\bc
\bt{c|c|c}\hline\hline
\TT \                   & {\it cons.}         & {\it viol.}           \\  \hline
\TT\BB $\pi\pi\to\pi\pi$   & $f_j^I $        &  $-$      \\  \hdashline
\TT\BB  $\eta\pi\to\eta\pi$ & $f^\etapi_j$      &  $-$     \\  \hdashline
\TT\BB  $\eta\pi\to\pi\pi$  & $-$             & ${\cal M}^I_j$ \\  \hdashline
\TT\BB  $\eta\pi\to K\Kbar$ & $g^\etapi_{j(even)}$ & ${\cal N}^0_{j(even)}\ ,
\ {\cal N}^1_{j(odd)}$ \\  \hdashline
\TT\BB  $ K\Kbar\to\pi\pi$  & $g^I_j$         & 
${\cal G}^{10}_{j(even)}\ ,\ {\cal G}^{12}_{j(even)}\ ,
\ {\cal G}^{01}_{j(odd)}$\\  \hdashline
\TT\BB  $K\Kbar\to K\Kbar$  & $h^I_j$         & ${\cal H}^{10}_j$  \\    \hline\hline
\et
\caption{\small Isospin-conserving (cons.) and
  isospin-violating (viol.) amplitudes involving $\pi\pi$, $\pi\eta$ and
  $K\Kbar$ channels.}
\lbltab{notations}
\ec
\end{table}
\section{Multichannel  Khuri-Treiman equations}
\subsection{Closed system of unitarity equations}
We can now write down a closed system of unitarity equations. It will
be convenient to introduce a matrix notation for the isospin
conserving amplitudes with $I=0$ and $I=1$
\be
\bm{T}^0=
\begin{pmatrix}
f_0^0 & g_0^0 \\
g_0^0 & h_0^0 \\
\end{pmatrix}\qquad
\bm{T}^1=
\begin{pmatrix}
f_0^{\eta\pi} & g_0^{\eta\pi} \\
g_0^{\eta\pi} & h_0^1 \\
\end{pmatrix}\qquad
\en
The $I=0$ amplitude ${\cal
  M}_0^0$ is now embedded into a system of four coupled unitarity equations
\be\lbl{coupledunit0}
\ba{ll}
\im\begin{pmatrix} 
{\cal M}_0^0 & {\cal G}^{10}_0\\
{\cal N}^0_0 & {\cal H}^{10}_0\\ \end{pmatrix}
= & {\bm{T}^0}^* \Sigma^0 
\begin{pmatrix} 
{\cal M}_0^0 & {\cal G}_0^{10}\\
{\cal N}^0_0 & {\cal H}^{10}_0\\ \end{pmatrix} 
 +\begin{pmatrix} 
{{\cal M}_0^0}^* & {{\cal G}_0^{10}}^*\\
{{\cal N}^0_0}^* & {{\cal H}^{10}_0}^*\\ 
\end{pmatrix} \Sigma^1 {\bm{T}^1} \\[0.3cm]
\  &+ {\bm{T}^0}^*  
\begin{pmatrix}
0 & 0\\
0 & \Delta\sigma_K \\
\end{pmatrix} \bm{T}^1
\ea\en
where
\be
\Sigma^0 =
\begin{pmatrix}
\sigma_\pi(s) & 0 \\
0            & \sigma_K(s)\\
\end{pmatrix}\qquad
\Sigma^1 =
\begin{pmatrix}
\sigma_{\eta\pi}(s) & 0 \\
0            & \sigma_K(s)\\
\end{pmatrix}\ 
\en
while the $I=2$ amplitude ${\cal M}_0^2$ is involved in a system of two
unitarity equations,
\be\lbl{coupledunit2}
\im\begin{pmatrix}
{\cal M}^2_0 \\[0.1cm]  
{\cal G}^{12}_0\end{pmatrix}
=\sigma_\pi (f_0^2)^*  
\begin{pmatrix}{\cal M}^2_0 \\[0.1cm] {\cal G}^{12}_0\end{pmatrix}
+\bm{T}^1\Sigma^1  \begin{pmatrix}{{\cal M}^2_0}^* \\[0.1cm] {{\cal G}^{12}_0}^*
\end{pmatrix}\ .
\en
Finally, for the $I=1$ amplitude ${\cal M}_1^1$, the coupled unitarity equations
read
\be
\im\begin{pmatrix}
{\cal M}_1^1\\
{\cal N}_1^1\\
\end{pmatrix}=
\begin{pmatrix}
f_1^1 & g_1^1\\
g_1^1 & h_1^1\\
\end{pmatrix}^* \Sigma^0
\begin{pmatrix}
{\cal M}_1^1\\
{\cal N}_1^1\\
\end{pmatrix}
\en
In the following, however, we will disregard the inelasticity effects
for ${\cal M}_1^1$ and continue to use the elastic unitarity
equation~\rf{unitrelelast} in this case. 

\subsection{Coupled Muskhelishvili-Omn\`es representation}
The next step is to write each one of the isospin violating partial-wave
amplitudes as a sum of two functions, one having a right-hand cut only and one
having a generalised left-hand cut. For physical mass values, the left and
right-hand cuts of the various partial-waves appear to be overlapping, but it
is possible to separate them unambiguously.  In the case of the amplitudes
involving the $\eta$ meson, this is done by using the $\metad+i\epsilon$
prescription. The amplitude $K\Kbar\to K\Kbar$ has a left-hand cut which
extends on the real axis in the range $[-\infty, 4\mkd-4\mpid]$. Using the
$\mkd+i\epsilon$ prescription shifts this cut above the real axis.

For the $I=0$ amplitudes one  writes
\be
\begin{pmatrix} 
{\cal M}_0^0(s) & {\cal G}_0^{10}(s)\\
{\cal N}^0_0(s) & {\cal H}_0^{10}(s)\\ \end{pmatrix}
=\dfrac{\sqrt6\epsilon_L}{32\pi}
\begin{pmatrix} 
 M_0(s) +\hat{M}_0(s) & G_{10}(s) +\hat{G}_{10}(s)\\
 N_0(s) +\hat{N}_0(s)&  H_{10}(s) +\hat{H}_{10}(s)\\ \end{pmatrix}
\en
which generalises eq.~\rf{calM00LR} while for the $I=2$ amplitudes one can
write 
\be
\begin{pmatrix}
{\cal M}_0^2(s) \\
{\cal G}_0^{12}(s)\\
\end{pmatrix}=
-\frac{\epsilon_L}{16\pi}
\begin{pmatrix}
M_2(s)+\hat{M}_2(s)\\
G_{12}(s)+\hat{G}_{12}(s)\\
\end{pmatrix}\ .
\en
One can now employ the standard Omn\`es method in
order to express the right-cut functions in terms of the left-cut
ones. Introducing the matrix notation
\be
\bm{M}_0(s)=
\begin{pmatrix}
M_0(s) & G_{10}(s)\\
N_0(s) & H_{10}(s)\\
\end{pmatrix},\quad
\hat{\bm{M}}_0(s)=
\begin{pmatrix}
\hat{M}_0(s) & \hat{G}_{10}(s)\\
\hat{N}_0(s) & \hat{H}_{10}(s)\\
\end{pmatrix}
\en
the discontinuity relation for the $\bm{M}_0$ functions is deduced from the
unitarity relation~\rf{coupledunit0},
\bea\lbl{discM0coupl}
&&\disc[\bm{M}_0(s)]= 
{\bm{T}^0}^*(s) \Sigma^0\,\big[\bm{M}_0(s+i\epsilon) +\hat{\bm{M}}_0(s)\big]
 \nonumber\\
&& \phantom{ \disc[\bm{M}_0(s)]  }
+\big[(\bm{M}_0(s-i\epsilon) +\hat{\bm{M}}_0(s)\big]\Sigma^1\,  {\bm{T}^1(s)}
+  {\bm{T}^0}^*(s) \Delta\Sigma_K {\bm{T}^1(s)} 
\ena
where
\be
\Delta\Sigma_K= {32\pi\over\sqrt6\,\epsilon_L}\,\Delta\sigma_K
\begin{pmatrix}
0 & 0 \\
0 & 1 \\
\end{pmatrix}\ .
\en
and $\Delta\sigma_K$ is given in eq.~\rf{Deltasig}. 
Eq.~\rf{discM0coupl} generalises the one-channel discontinuity
relation~\rf{discMI}. 

Let us now consider the following matrix
\be\lbl{Xdef}
\bm{X}(s)= \bm{\Omega}_0^{-1}(s) \bm{M}_0(s) 
{}^t\bm{\Omega}_1^{-1}(s)
\en
where $\bm{\Omega}_I$ are the $2\times2$ Muskhelishvili-Omn\`es matrices
corresponding to the $T$-matrices $\bm{T}_I$. Making use of the following
discontinuity properties of the MO matrices
\bea\lbl{MOdisc}
&&\bm{\Omega}_0(s+i\epsilon)= (1+2i \bm{T}^0\Sigma^0)\bm{\Omega}_0(s-i\epsilon)
=(1-2i {\bm{T}^{0}}^*\Sigma^0)^{-1} \bm{\Omega}_0(s-i\epsilon)\nonumber\\
&&
{}^t\bm{\Omega}_1(s+i\epsilon)= 
{}^t\bm{\Omega}_1(s-i\epsilon)(1+2i \Sigma^1\bm{T}^1)
={}^t\bm{\Omega}_1(s-i\epsilon)(1-2i \Sigma^1{\bm{T}^1}^*)^{-1} 
\ena
one can express the discontinuity of the $\bm{X}$ matrix elements in terms of
the $\hat{\bm{M}}_0$ functions and $\Delta\sigma_K$
\be
\disc[\bm{X}(s)] = \Delta\bm{X}_a(s) + \Delta\bm{X}_b(s)
\en
where
\be\lbl{discXa}
\Delta\bm{X}_a=
\bm{\Omega}^{-1}_0(s-i\epsilon) \left[  
{\bm{T}^0}^*(s) \Sigma^0 \hat{\bm{M}}_0(s+i\epsilon)
+\hat{\bm{M}}_0(s-i\epsilon) \Sigma^1 {\bm{T}^1}(s)
\right] {}^t\bm{\Omega}^{-1}_1(s+i\epsilon)
\en
An alternative expression for $\Delta\bm{X}_a$ can be derived, using
eqs.~\rf{MOdisc})  
\be\lbl{discX}
\ba{l}
\Delta\bm{X}_a= \\
-\left\{\im [\bm{\Omega}^{-1}_0(s+i\epsilon)]\,\hat{\bm{M}}_0(s)\,
{}^t\bm{\Omega}^{-1}_1(s+i\epsilon)
+\bm{\Omega}^{-1}_0(s-i\epsilon)\, \hat{\bm{M}}_0(s)\,
\im[ {}^t\bm{\Omega}^{-1}_1(s+i\epsilon)]  \right\}        \\
\ea\en
which shows that it represents the discontinuity of the following quantity 
\be
\Delta\bm{X}_a=-\disc[\bm{\Omega}^{-1}_0(s) \,\hat{\bm{M}}_0(s)\, 
{}^t\bm{\Omega}^{-1}_1(s)]
\en
across the right-hand cut.
The quantity $\Delta\bm{X}_b$ is proportional to $\Delta\sigma_K$ and it is
given by 
\be\lbl{discXb}
\Delta\bm{X}_b={32\pi\over\sqrt6\,\epsilon_L}
\Delta\sigma_K \,\bm{\Omega}^{-1}_0(s-i\epsilon) 
{\bm{T}^0}^*(s) \begin{pmatrix}
0&0\\
0&1\\ \end{pmatrix}{\bm{T}^1}(s)
 {}^t\bm{\Omega}^{-1}_1(s+i\epsilon)\ .
\en
We can then write a twice subtracted dispersive representation for $\bm{X}(s)$
and generate, via~\rf{Xdef}, a MO representation for the $\bm{M}_0$
amplitudes 
\be\lbl{M0disp}
\begin{pmatrix}
M_0(w) & G_{10}(w)\\
N_0(w) & H_{10}(w)\\
\end{pmatrix}= \bm{\Omega}_0(w)\left[ \bm{P}_0(w)
+ w^2\,(\hat{\bm{I}}_{a}(w) +  \hat{\bm{I}}_{b}(w))
\right]{}^t\bm{\Omega}_1(w)
\en
where $\bm{P}_0$ is a $2\times2$ matrix of polynomial
functions,
\be
\bm{P}_0(w)=\begin{pmatrix}
\alpha_0 +\beta_0 w + \gamma_0 w^2 & 
\alpha^G_0 +\beta^G_0 w + \gamma^G_0 w^2\\
\alpha^N_0 +\beta^N_0 w + \gamma^N_0 w^2 & 
\alpha^H_0 +\beta^H_0 w + \gamma^H_0 w^2\\
\end{pmatrix}
\en
and the integral parts are
\be\lbl{Ihata,b}
\hat{\bm{I}}_{a,b}=
{1\over\pi}\int_{4\mpid}^\infty {ds'\over (s')^2 (s'-w)} 
\,\Delta\bm{X}_{a,b}(s')\ .
\en
One remarks that in the term $\Delta\bm{X}_b$ the quark mass ratio
$\epsilon_L$ appears in the denominator and thus cancels with the overall
factor in the complete amplitudes. This part is driven by the physical
$K^+-K^0$ mass difference via the $\Delta\sigma_K$ function
(see~\rf{Deltasig}). This mechanism was first studied in
ref.~\cite{Achasov:1979xc} who predicted that a large isospin violation should
take place at 1 GeV in the $\eta\pi\to \pi\pi$ scattering amplitude, due to
the contributions from both the $a_0(980)$ and the $f_0(980)$ resonances. The
set of eqs.~\rf{M0disp} account for the other sources of isospin
violation as well.

We now consider the case of the $I=2$ amplitudes and we also introduce a
matrix notation for the column matrices
\be
\bm{M}_2= \begin{pmatrix}
M_2(s)\\
G_{12}(s)\\
\end{pmatrix},\quad
\hat{\bm{M}}_2= \begin{pmatrix}
\hat{M}_2(s)\\
\hat{G}_{12}(s)\\
\end{pmatrix}
\en
From the unitarity relations including the $\pi\pi$, $\pi\eta$ and $K\Kbar$
contributions we deduce the discontinuity of the $\bm{M}_2$ functions
\be\lbl{discM2}
\disc[\bm{M}_2(s)]=
\sigma_\pi(f_0^2(s))^* (\bm{M}_2(s+i\epsilon)+\hat{\bm{M}}_2(s))
+ \bm{T}^1(s)\Sigma^1(\bm{M}_2(s-i\epsilon)+\hat{\bm{M}}_2(s))\ .
\en 
As before, we introduce a matrix $\bm{X}_2$, multiplying $\bm{M}_2$ by inverse
MO functions,
\be
\bm{X}_2=\Omega_2^{-1}\, \bm{\Omega}_1^{-1} \bm{M}_2\ .
\en
Its discontinuity relation is expressed in terms of the $\hat{\bm{M}}_2$
functions and the MO functions
\be\lbl{discX2one}
\Delta \bm{X}_2
=\Omega_2^{-1}(s-i\epsilon)\bm{\Omega}_1^{-1}(s+i\epsilon)\left[ 
\sigma_\pi {f_0^{2}}^*(s)\hat{\bm{M}}_2(s)
+ \bm{T}^1(s)\Sigma^1\hat{\bm{M}}_2(s)\right]\ .
\en
An alternative useful expression for $\Delta\bm{X}_2$ can be derived
\begin{align}\lbl{altDX2}
\Delta\bm{X}_2 = & -\Big\{
\im[ \Omega_2^{-1}(s+i\epsilon) ] \bm{\Omega}_1^{-1}(s+i\epsilon)
 +\Omega_2^{-1}(s-i\epsilon)\,\im [\bm{\Omega}_1^{-1}(s+i\epsilon)]
\Big\}\hat{\bm{M}}_2(s) \nonumber \\
\ = & -\disc\big[ 
\Omega_2^{-1}(s) \bm{\Omega}_1^{-1}(s)\big]\, \hat{\bm{M}}_2(s)\ .
\end{align}
This leads to the following integral MO representation for the 
matrix $\bm{M}_2$,
\begin{align}\lbl{M2disp}
 \begin{pmatrix}
M_2(w)\\
G_{12}(w)\\
\end{pmatrix}= &
\Omega_2(w)\bm{\Omega}_1(w)
\begin{pmatrix}
          \hat{I}_2(w)\,w^2\\
\alpha_2^K+\beta_2^K\,w +\left(\gamma_2^K+\hat{I}_2^K(w)\right)\,w^2\\
\end{pmatrix}
\end{align}
with
\be
 \begin{pmatrix}
\hat{I}_2(w)\\
\hat{I}_2^K(w)\\
\end{pmatrix}= 
 {1\over\pi}
\int_{4\mpid}^\infty {ds'\over (s')^2(s'-w)}\,\Delta\bm{X}_2(s')\ .
\en

Eqs.~\rf{M0disp} and ~\rf{M2disp} together with the uncoupled equation for
$M_1$~\rf{dispomnes} involve 16 polynomial parameters. The polynomial
dependence was chosen such that the equations would reduce exactly to the set
of elastic equations~\rf{dispomnes} if one switches off the coupling to the
$K\Kbar$ channel as well as $\eta\pi$ rescattering\footnote{The one-channel
  case is recovered by setting $(\bm{T}^1)_{ij}=0$ and the MO matrices to
  $[\bm{\Omega}_0]_{ij}=\delta_{ij}(1+\delta_{i1}\Omega_0)$ and
  $[\bm{\Omega}_1]_{ij}=\delta_{ij}$.}. One remarks that the asymptotic
behaviour in the coupled-channel equations is modified as compared to the
one-channel case: since the matrix elements of the MO matrices
$\bm{\Omega}_I$, decrease as $1/s$ when $s$ goes to $\infty$ the entries of
the $\bm{M}_0$ matrix (thus the $M_0$ function) behave as constants. The
asymptotic behaviour of $M_2$, in contrast, remains the same as before.

In addition to the polynomial parameters, the right-hand side of
eqs.~\rf{M0disp},~\rf{M2disp} involve a number of ``hat functions''. The
functions $\hat{M}_I$ are determined in terms of the functions $M_I$ by the
angular integrals~\rf{Mhatexpr}, but one still needs to determine the other
hat functions $\hat{G}_{10}$, $\hat{N}_0$, $\hat{H}_{10}$, $\hat{G}_{12}$
which are related to the $K\bar{K}$ amplitudes.  For this purpose, one would
have to consider all the related crossed channel amplitudes an write similar
sets of equations (which would introduce further one-variable
functions). Here, since we are mainly interested in the $\eta\pi\to \pi\pi$
amplitude we will content with an approximation for the amplitudes involving
the $K\Kbar$ channel, simply neglecting the integrals which involve the
left-cut functions, i.e. we take
\be
\hat{G}_{10}= \hat{N}_0= \hat{H}_{10}= \hat{G}_{12}=0. 
\en
In support of this approximation, one observes that if one were to neglect the
left-cut integrals in the $\eta\to 3\pi$ amplitude itself, one would still
obtain a qualitatively reasonable description (e.g.~\cite{Bijnens:2002qy}). 
With this approximation, eqs.~\rf{M0disp} and ~\rf{M2disp} constitute a closed
set of equations which form a coupled-channel generalisation of the
Khuri-Treiman equations~\rf{dispomnes} for $M_0$ and $M_2$.
 
\subsection{Matching to the chiral amplitudes}
We intend to fix the 16 polynomial parameters by matching to the chiral
expansions of the amplitudes involved. For the $\eta \to 3\pi$ amplitude we
will use the NLO expansion also including the part of the electromagnetic
contributions of order $e^2(m_d+m_u)$ from ref.~\cite{Baur:1995gc}.  This
makes it possible to display explicitly the term induced by the $K^+-K^0$ mass
difference via unitarity which, in the dispersive representation, is contained
in the $\hat{\bm{I}}_b$ integrals (see eqs.~\rf{discXb},~\rf{Ihata,b}). The
explicit chiral expressions for the functions $M_I$ as used here are given in
appendix~\ref{sec:appendixA}. 

For the isospin violating amplitudes involving the $K\Kbar$ channel we will
use the leading order chiral expansion.  At this order, the partial-wave
amplitudes have no left-hand cut, which is consistent with the approximation
of dropping the left-cut functions in the integral equations. The relevant
expressions (including the $O(e^2)$ contributions) are given below
\begin{align}\lbl{amplitKKp2}
\bar{G}_{10}(w)= & \frac{\sqrt6}{\Delta_{K\pi}}\,\left(
\frac{3}{8}\,w- \frac{1}{2} \mkd \right) 
+\frac{2\sqrt2}{3\,\epsilon_L} \frac{e^2 C}{\fpiq} \nonumber\\
\bar{N}_0(w)= & \frac{\sqrt3}{\Delta_{K\pi}}\,\left(
-\frac{3}{4}\,w+ \mpid \right) \nonumber\\
\bar{H}_{10}(w)= & \frac{4}{\sqrt6\,\epsilon_L} \frac{e^2 C}{\fpiq} \nonumber\\
\bar{G}_{12}(w)= & \frac{\sqrt6}{\Delta_{K\pi}}\,\left(
\frac{9}{16}\,w-\frac{3}{4}\mkd \right) 
-\frac{1}{\sqrt2\,\epsilon_L} \frac{e^2 C}{\fpiq}
\end{align}
with
\be
\Delta_{PQ}= m^2_P- m^2_Q
\en
and the coupling constant $C$ can be related to the $\pi^+-\pi^0$ mass
difference (see appendix~\ref{sec:appendixA}).

We implement matching conditions for the $\eta\to 3\pi$ amplitude following
the simple method of ref.~\cite{Descotes-Genon:2014tla} which differs slightly
from that of ref.~\cite{Anisovich:1996tx}. 
Let $\bar{M}(s,t,u)$ be the amplitude computed in the chiral expansion at
order $p^4$. The polynomial parameters of the dispersive amplitude 
must be determined such that the dispersive and chiral amplitudes coincide
for small values of the Mandelstam variables. At order $p^4$ one
should thus have
\be\lbl{genericmatch}
M(s,t,u)- \bar{M}(s,t,u) =O(p^6)\ .
\en
This condition is satisfied automatically for the discontinuities, which
implies that one can neglect the discontinuity of the differences of the
one-variable functions $M_I-\bar{M}_I$ and thus expand these differences as
polynomials,
\be
M_I(w)- \bar{M}_I(w) = \sum_{n=0}^{n_I} \lambda_n\, w^n
\en
(with $n_0=n_2=2$, $n_1=1$). Inserting these expansions into the amplitude
difference $M(s,t,u)- \bar{M}(s,t,u)$  and requiring that the $O(p^4)$
polynomial part vanishes gives four equations. In the elastic Khuri-Treiman
framework these determine the four parameters via the following four equations
\begin{align}\lbl{matcheq1} 
\alpha_0= & \bar{M}_0(0)+\frac{4}{3}\,\bar{M}_2(0)
           +3s_0\,( \bar{M}'_2(0)-  \bar{M}_1(0))
           +9s_0^2\,\bar{M}_2^{eff} \nonumber\\
\beta_0= & \bar{M}'_0(0)+3\,\bar{M}_1(0)-\frac{5}{3}\,\bar{M}'_2(0)
           -9s_0\,\bar{M}_2^{eff} -\Omega'_0(0)\,\alpha_0 \nonumber\\
\beta_1= & \bar{M}'_1(0)-\hat{I}_1(0) + \bar{M}_2^{eff}\nonumber\\
\gamma_0= & \frac{1}{2}\,\bar{M}''_0(0)- \hat{I}_0(0)+\frac{4}{3}\,\bar{M}_2^{eff} 
-\frac{1}{2} \Omega''_0(0) \,\alpha_0 - \Omega'_0(0)\,\beta_0
\end{align} 
with
\be
\bar{M}_2^{eff}=\frac{1}{2}\,\bar{M}''_2(0)-\hat{I}_2(0)\ 
\en
and one must keep in mind that the integrals $\hat{I}_I(0)$ which appear on
the right-hand sides depend linearly on the four parameters. In
ref.~\cite{Anisovich:1996tx} the first two of eqs.~\rf{matcheq1} were replaced
by two equations related to the position $s_A$ of the Adler zero of the
chiral amplitude $\bar{M}(s,t,u)$ along the line $u=s$ and the value of its
derivative at $s=s_A$. We will see below that eqs.~\rf{matcheq1} do
actually implement these Adler zero conditions to a rather good
approximation. Additionally, approximations were
made in ref.~\cite{Anisovich:1996tx} in the determination of $\beta_1$ and
$\gamma_0$ (yielding e.g. $\gamma_0\simeq 0$), the validity of which depends on
the assumed behaviour of the phase $\delta^0$ in the inelastic region. Here,
the four equations will be solved without approximation. Doing so, one notes
that the polynomial parameters get an imaginary part from the contributions of
the integrals $\hat{I}_I(0)$ which, however, is small (less that 10\% of the
real part).

In the coupled-channel case, the matching of the $\eta\to 3\pi$ amplitude
again gives rise to four equations. In addition, we can match the values at
$w=0$ of each one of the four $K\Kbar$ amplitudes with the chiral ones given
in eqs.~\rf{amplitKKp2} as well as the values of the first and second
derivatives (which are vanishing). Altogether, this provides 16 constraints
which fix all the polynomial parameters appearing in the coupled-channel
Khuri-Treiman equations. The details of these matching equations are given in
appendix~\ref{sec:appendixB}.

\section{Results and comparisons with experiment}
\subsection{Numerical method}
The main difficulty which is involved in deriving accurate numerical solutions
of the Khuri-Treiman equations is tied to the evaluation of the angular
integrals $\braque{z^n M_I}$ needed to obtain the hat functions $\hat{M}_I$
and to the treatment of the singularities of these functions in the
computation of the $\hat{I}_a$ integrals which are finite.  These technical
aspects are discussed in detail in ref.~\cite{Kambor:1995yc}. By a change of
variables one can rewrite the angular integrals as integrals over $t$
\be\lbl{znMIint}
\braque{z^n M_I}(s )={1\over (\kappa(s))^{n+1}}
\int_{t^-(s)}^{t^+(s)} (2t+s-3s_0)^n\,M_I(t)\,dt ,
\en
with
\be\lbl{t+-}
t^\pm(s)={1\over2}( 3s_0 -s  \pm \kappa(s) )\ .
\en
When $s$ lies in the range $(\meta-\mpi)^2 < s < (\meta+\mpi)^2$, the
endpoints $t^\pm(s)$ become complex. In fact, using the $\metad+i\epsilon$
prescription one sees that $t^+$ and $t^-$ are placed on opposite sides of the
unitarity cut when $s$ gets larger that $(\metad-\mpid)/2$.  The integral from
$t^-$ to $t^+$ must then be evaluated along a complex contour which circles
around the unitarity cut of the functions $M_I$. Rather than computing
explicitly $M_I(w)$ for complex values of $w$, as in
ref.~\cite{Kambor:1995yc}, we follow here the approach of
ref.~\cite{Descotes-Genon:2014tla} which consists in inserting the dispersive
representations~\rf{disprelMI} of the $M_I$ functions into eq.~\rf{znMIint}
and computing analytically the $t$ integrals. This makes it possible to express
the functions $\braque{z^n M_I}$ in terms of the discontinuities $\disc[M_I]$
along the positive real axis. The relevant formulae are recalled in
appendix~\ref{sec:appendixC}. 

The equations are conveniently solved using an iterative procedure. On the
first iteration step, one sets the $\hat{M}_I$ functions equal to zero. Then
the $\hat{I}$ integrals are also equal to zero and it is straightforward to
compute the values of the polynomial parameters from the matching equations
and then the  functions $M_I$, $G_{10}$, $N_0$, $H_{10}$, $G_{12}$ as well as
the discontinuities $\disc[M_I]$ (which are given from eqs.~\rf{discMI} in the
one-channel case and ~\rf{discM0coupl}~\rf{discM2} in the coupled-channel case).
The coupled-channel framework is somewhat more complicated to handle than the
single-channel one, essentially because the MO matrices do not
obey a simple explicit representation in terms of the $T$-matrix elements and
must be solved for numerically, but does not otherwise involve any specific
difficulty. Then, on each iteration step, one updates the values of the
functions $\hat{M}_I$ using $\disc[M_I]$ from the preceding step, and then
compute the $\hat{I}_a$ integrals. Then, one updates the values of the
polynomial parameters and derive the new values of all the functions $M_I$,
$G_{10}$, $N_0$, $H_{10}$, $G_{12}$ and those of discontinuities
$\disc[M_I]$. 

Convergence is found to be reasonably fast. Denoting by $M_I^{(n)}$ the result
obtained at the $n^{th}$ iteration step we can estimate the rate of
convergence from the quantity
\be
\epsilon^{(n)} =\max_{I,s}\left\vert
\frac{M_I^{(n)}(s)- M_I^{(n-1)}(s)  }{M_I^{(n)}(s) }
\right\vert\ .
\en
Anticipating on the numerical results to be presented below, with $n=5,6,7$ we
find: $\epsilon^{(5)}\simeq 4\cdot 10^{-3}$, $\epsilon^{(6)}\simeq 2\cdot
10^{-4}$,  $\epsilon^{(7)}\simeq 4\cdot 10^{-5}$. The $I=2$ amplitude is the
one which has the slowest convergence rate.

\begin{figure}[h]
\centering
\includegraphics[width=0.60\linewidth]{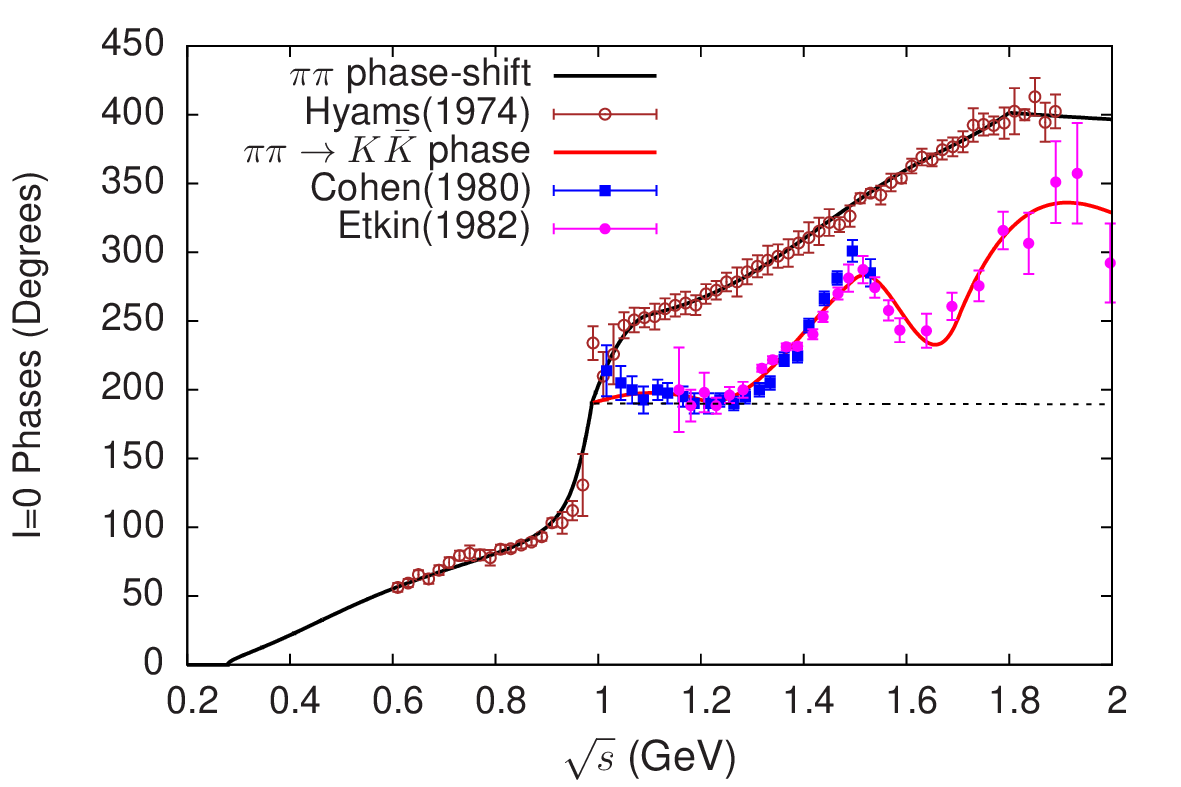}
\includegraphics[width=0.60\linewidth]{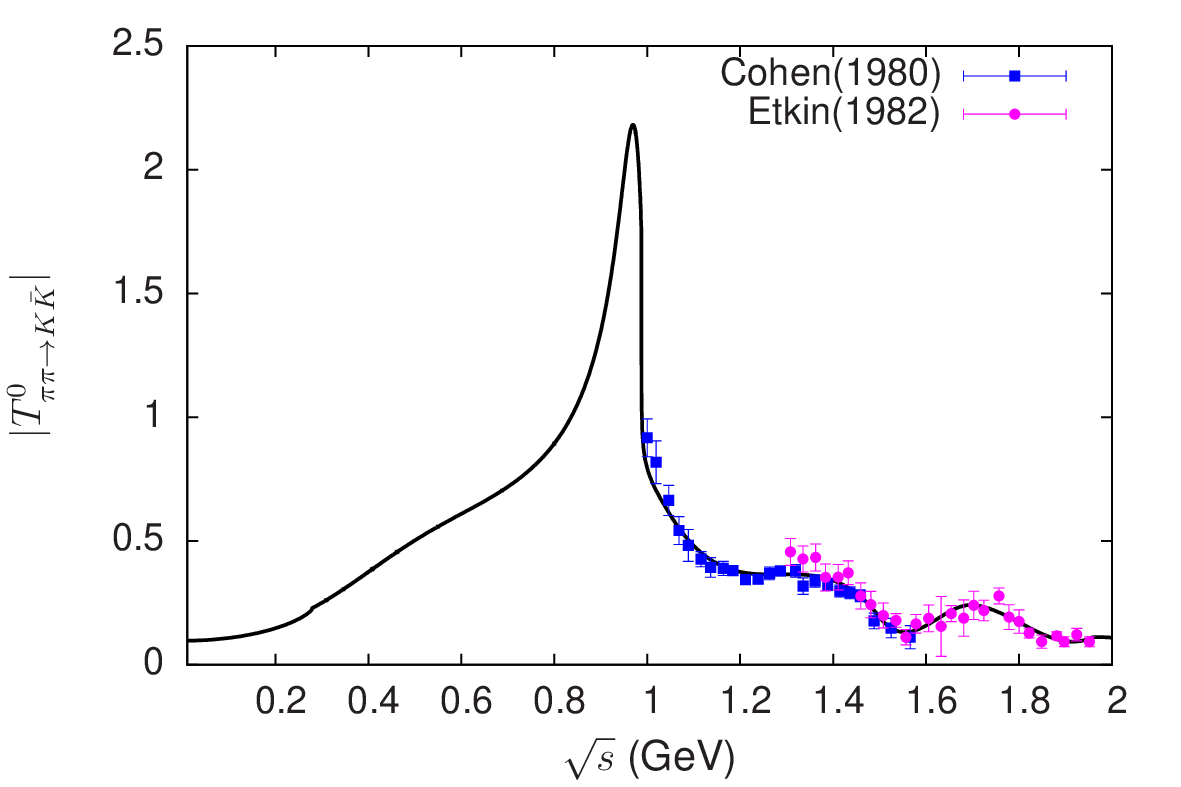}
\caption{\small Illustration of the $I=0$ $T$-matrix used. The upper figure
  shows the $\pi\pi$ phase-shift (upper curve) and the phase of the $\pi\pi\to
  K\Kbar$ amplitude $g_0^0$ (lower curve). The dotted line is the phase used
  for computing the one-channel Omn\`es function.  The lower figure shows the
  modulus of $g_0^0$. The curve is a fit to the data
  above the $K\Kbar$ threshold and an extrapolation, based on analyticity,
  below.}
\label{fig:I0inputs}
\end{figure}
\subsection{Input $I=0,1$  T-matrices}
Above the $K\bar{K}$ threshold, the $S$ and $T$ matrices are related by 
\be
\bm{S}= \bm{1} +2i \sqrt{\Sigma}\, \bm{T} \sqrt{\Sigma} 
\en 
Two-channel unitarity implies that all the $T$-matrix elements are determined
from three real inputs: a) the phase of $S_{11}$, b) the modulus of $T_{12}$ and
c) the phase of $T_{12}$.  For $I=0$ scattering, experimental data exist for
these quantities up to 2 GeV, approximately.  We will use here a determination
based on the experimental data of Hyams et al.~\cite{Hyams:1973zf} for the
$\pi\pi\to \pi\pi$ phase-shifts and the data of refs.~\cite{Cohen:1980cq}
and~\cite{Etkin:1981sg} (above 1.3 GeV) for the $\pi\pi\to K\Kbar$
amplitude\footnote{We note, however, that a recent analysis of the $\pi\pi$
  Roy equations in a once-subtracted version\cite{GarciaMartin:2011cn} shows
  some tension with the data of Cohen et al.~\cite{Cohen:1980cq} assuming that
  it saturates $\pi\pi$ inelasticity near the $K\Kbar$ threshold.}.  In the
higher energy region, a smooth interpolation is performed with the following
asymptotic conditions,
\be
\lim_{s\to\infty} \delta^0_{\pi\pi\to K\Kbar}(s) =2\pi,\quad
\lim_{s\to\infty} |T^0_{\pi\pi\to K\Kbar}(s)\vert =0
\en
which insure (in general) the existence of a corresponding unique MO
matrix~\cite{Muskhelishvili}. Below the $K\Kbar$ threshold, a determination of
the phase-shift based on the data of ref.~\cite{Hyams:1973zf} together with
constraints from the $\pi\pi$ Roy equations (in the twice-subtracted version
of ref.~\cite{Ananthanarayan:2000ht}) is performed.  It is assumed that
inelasticity can be neglected in this region, which implies that the phase of
the $\pi\pi\to K\Kbar$ amplitude coincides with the $\pi\pi$ phase-shift
(Fermi-Watson theorem). This
allows one to determine the modulus of this amplitude below the $K\Kbar$
threshold~\cite{Frazer:1960zza} where it is also needed. Details on the
parametrisation can be found in
refs.~\cite{GarciaMartin:2010cw,Moussallam:2011zg}.  Fig.~\fig{I0inputs} shows
the experimental data used and the fitted curves.

\begin{figure}[ht]
\centering
\includegraphics[width=0.60\linewidth]{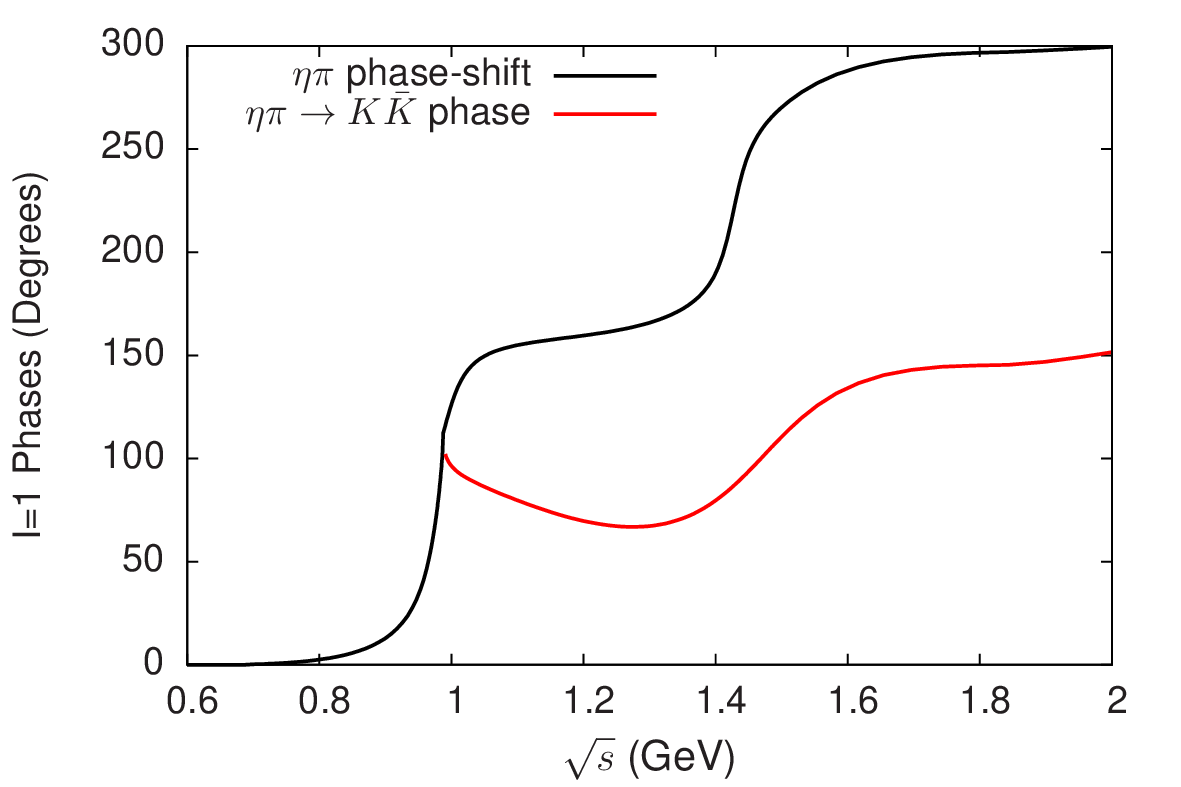}
\includegraphics[width=0.60\linewidth]{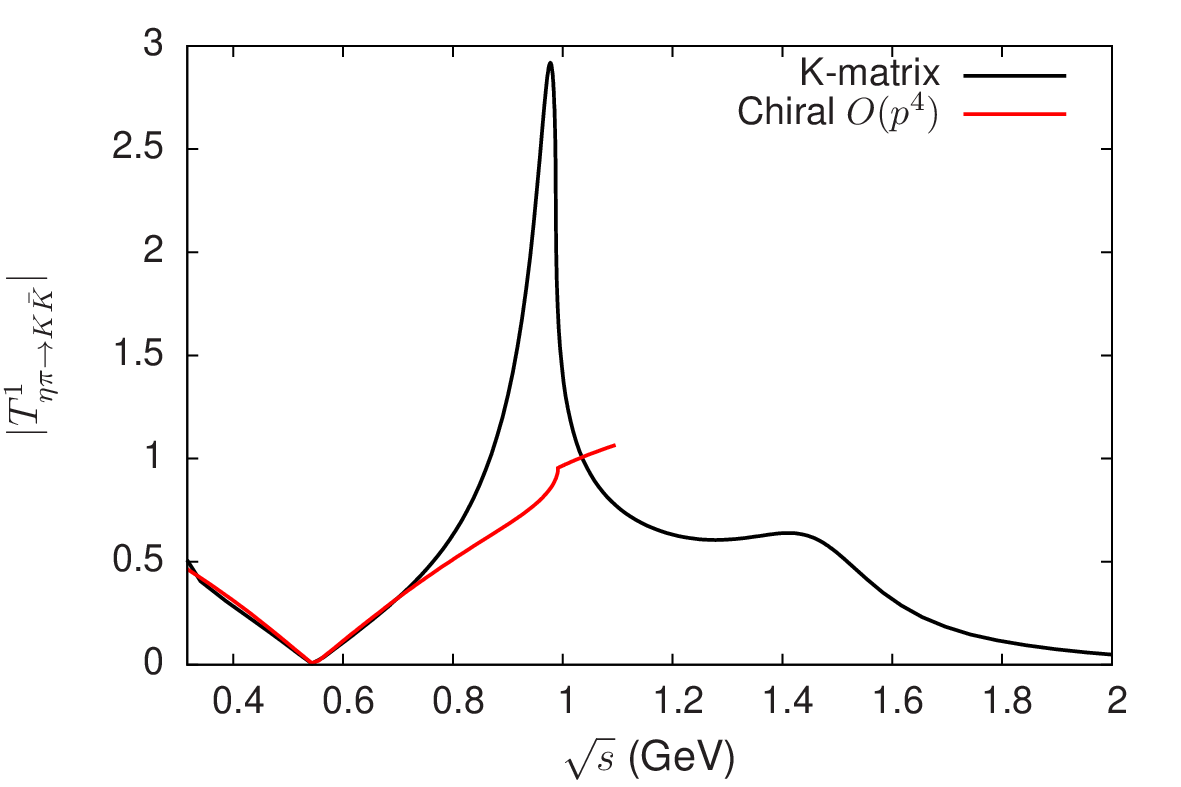}
\caption{\small Illustration of the $I=1$ $T$-matrix. The upper figure
  shows the $\eta\pi$ phase-shift and the phase of the $\eta\pi\to
  K\Kbar$ amplitude $g_0^{\eta\pi}$ (minus $\pi$). The lower figure
  shows the modulus of $g_0^{\eta\pi}$ which is also compared with the
  chiral $O(p^4)$ result.}
\label{fig:I1inputs}
\end{figure}
In the case of the $I=1$ $T$-matrix, experimental information on $\eta\pi$ and
$K\Kbar$ scattering are indirect and far less detailed than for $I=0$.  We
will rely here on the chiral $K$-matrix model proposed in
ref.~\cite{Albaladejo:2015aca} which provides a simple interpolation between
certain known properties of the prominent $I=1$ scalar resonances $a_0(980)$
and $a_0(1450)$ and the low energy properties constrained by chiral
symmetry. The $T$-matrix reproduces, by construction, the $\eta\pi\to \eta\pi$
and $\eta\pi\to K\Kbar$  chiral amplitudes at NLO (and more approximately the
amplitude $K\Kbar \to K\Kbar$). 
It depends on the values of the $O(p^4)$ chiral couplings: we use here (as in
ref.~\cite{Albaladejo:2015aca}) a set of values for these taken from a $p^4$
fit of ref.~\cite{Bijnens:2014lea}. We note that the value of $L_3$ in this
set is $L_3=-3.82\cdot10^{-3}$. We will use this value also in the computation
of the $\eta\to 3\pi$ amplitude, for consistency.

A further constraint can be implemented by computing the $\eta\pi$ and
$K\Kbar$ scalar form-factors from this $T$-matrix. Chiral symmetry relates the
$\eta\pi$ and the $K\pi$ scalar radii, which leads to the prediction that
$\braque{r^2}_S^{\eta\pi}$ should be remarkably small,
$\braque{r^2}_S^{\eta\pi}\simeq 0.1$ $\hbox{fm}^2$.  This small value can be
reproduced provided the phase $\delta_{\eta\pi\to K\Kbar}$ raises sufficiently
slowly above the $K\Kbar$ threshold.  The phenomenological parameters of the
model are also constrained by the properties of the
resonances. Fig.~\fig{I1inputs} shows a typical result for the $\eta\pi$ phase
shift and for the phase and modulus of the $\eta\pi\to K\Kbar$
amplitude\footnote{The phase shown is that of $T_0(\eta\pi^+ \to K^+\Kzb)=
  -g_0^{\eta\pi}$ (according to the isospin convention of
  eq.~\rf{isoassign}). It satisfies $\delta^0_{\eta\pi\to
    K\Kbar}(m_{a_0(1450)})=100\degree$ which corresponds to
  $\braque{r^2}^{\eta\pi}_S= 0.12$ $\hbox{fm}^2$, within 20\% of the chiral
  $O(p^4)$ result.}  which we will use in the present work.

\begin{figure}[th]
\centering
\includegraphics[width=0.60\linewidth]{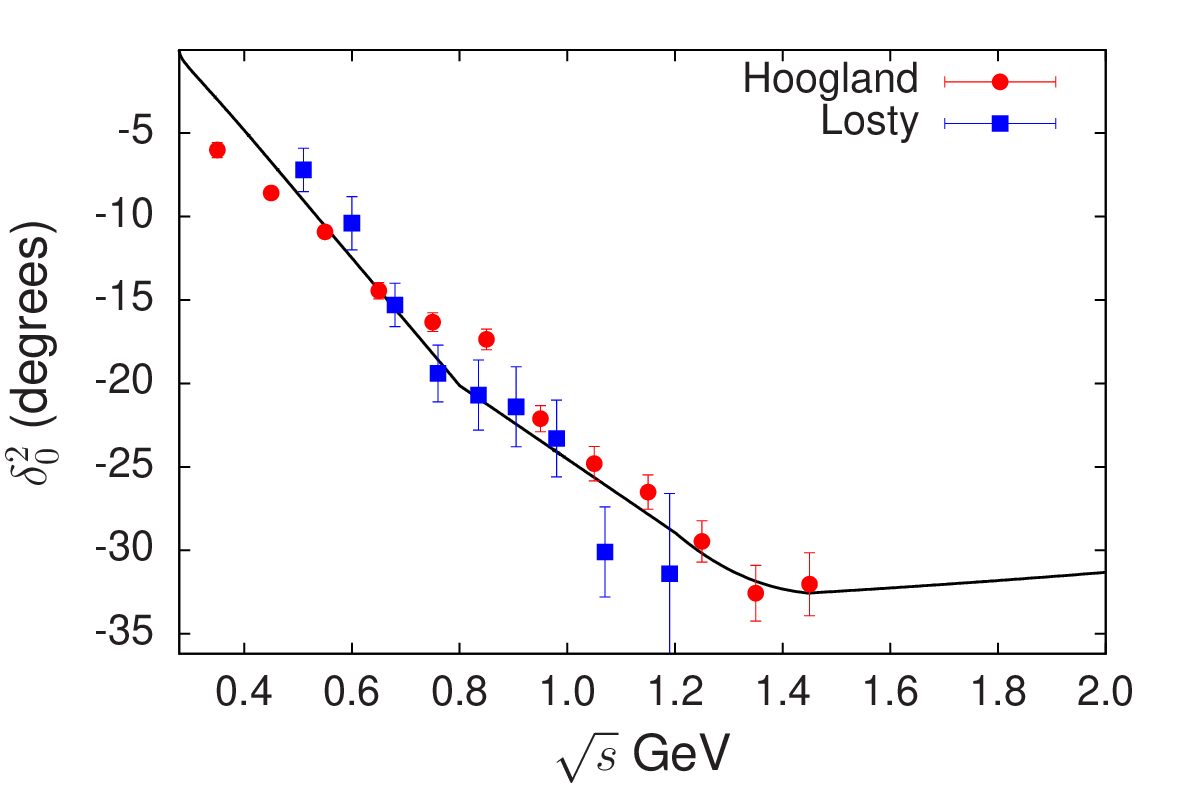}\\
\includegraphics[width=0.60\linewidth]{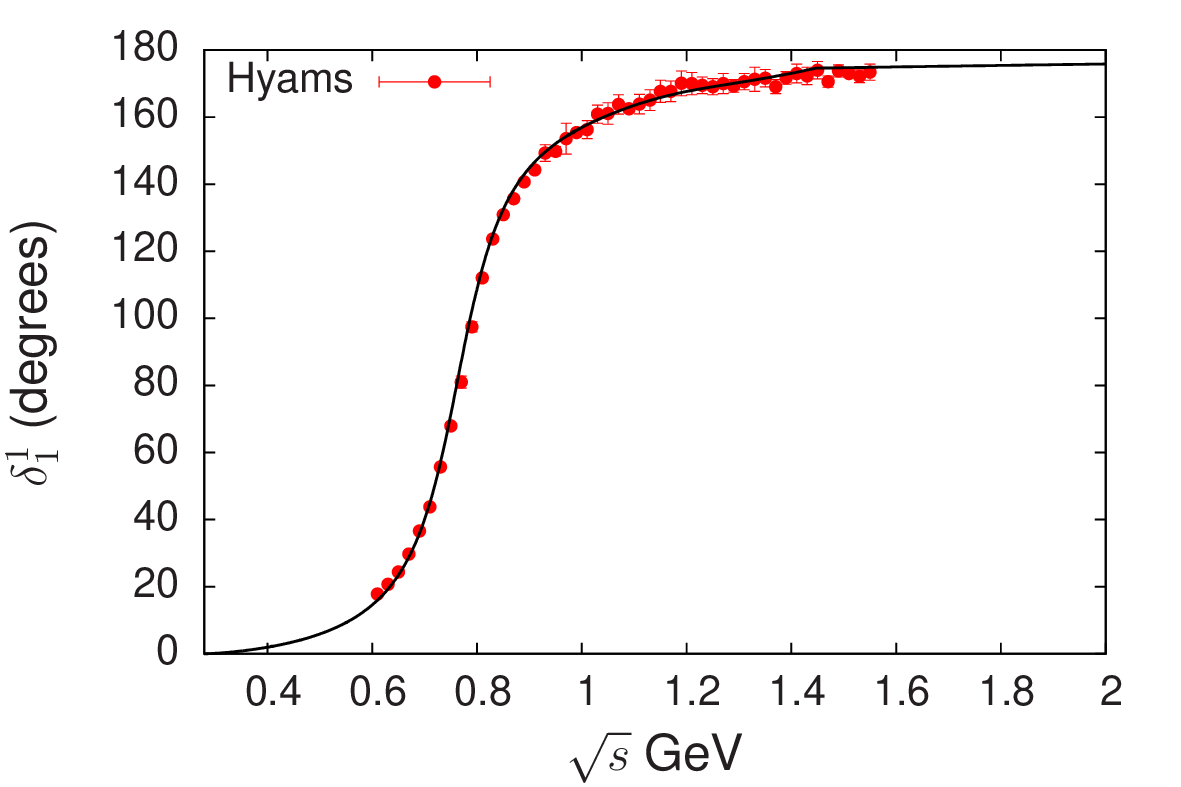}
\caption{\small The $\pi\pi$ phase shifts used for computing single-channel
  Omn\`es functions for the $I=2$ $S$-wave and the $I=1$ $P$-wave.}
\label{fig:I=1,2}
\end{figure}
Finally, the phase shifts which we used for the $I=1$ $\pi\pi$ $P$-wave and
the $I=2$ $S$-wave, for which inelasticity is ignored, are shown in
fig.~\fig{I=1,2}. In the energy region $\sqrt{s}\le 0.8$ GeV, these
phase shifts are given by the Roy equations solution parametrisation
of ref.~\cite{Ananthanarayan:2000ht}. They are fitted to experimental
data from~\cite{Hyams:1973zf} ($P$-wave)
and~\cite{Hoogland:1977kt,Losty:1973et} ($I=2$) in the region
  $\sqrt{s}\le 1.5$ GeV and interpolated to $\delta_1^1(\infty)=\pi$,
  $\delta_0^2(\infty)=0$ in the higher energy region.

\subsection{Illustration of the role of the inelastic channels}\label{sec:inelrole}
\begin{figure}[h]
\centering
\includegraphics[width=0.60\linewidth]{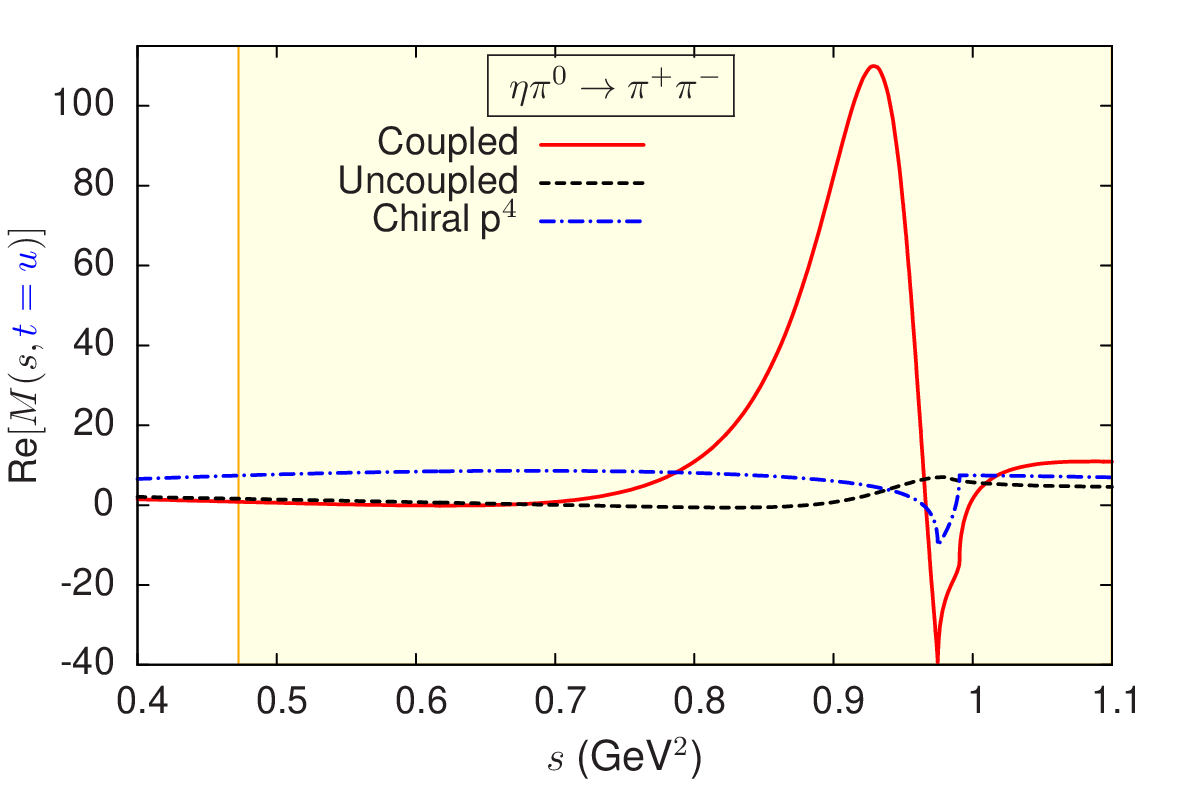}
\includegraphics[width=0.60\linewidth]{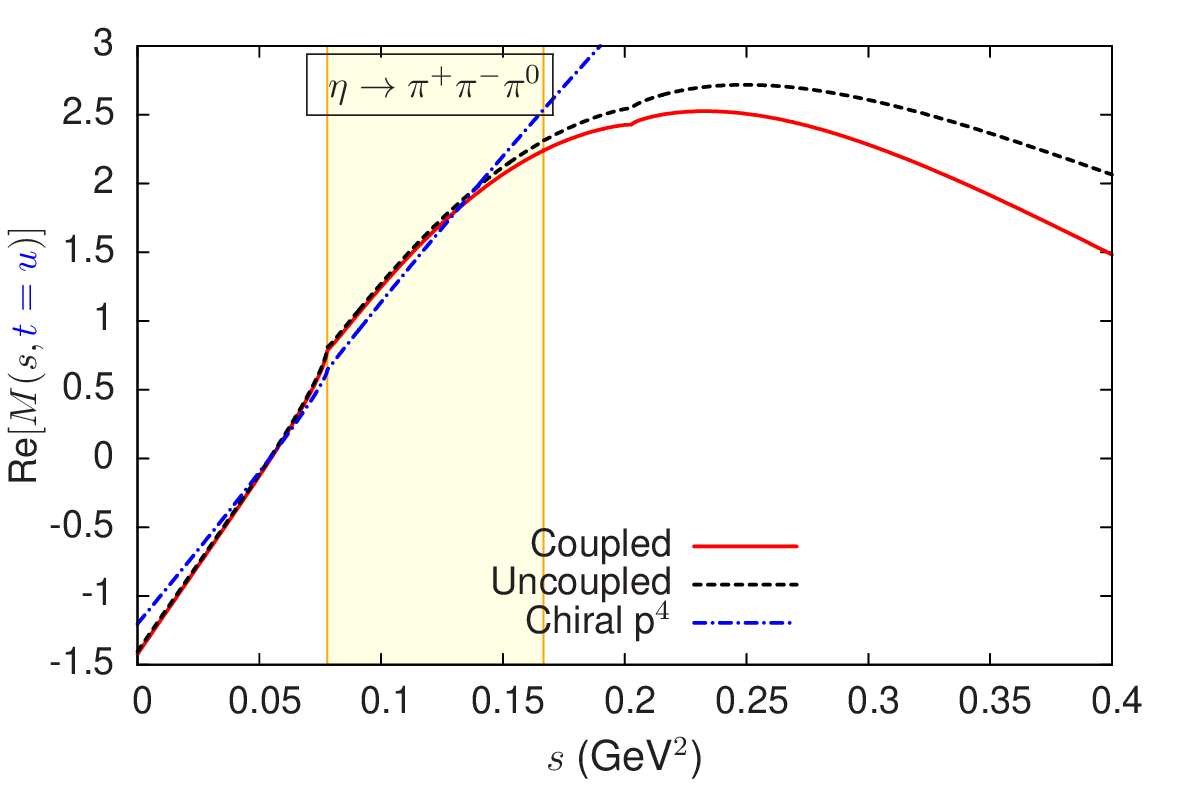}
\caption{\small Real part of the $\eta\to 3\pi$ amplitude along the $t=u$ line
  as a function of the energy $s$ in two different regions.  The yellow shaded
  area indicate the physical regions for the scattering $\eta\pi^0\to
  \pip\pim$ (upper figure) and the the decay $\eta\to \pip\pim\pi^0 $ (lower
  figure). The red solid curve corresponds to the solution of the
  coupled-channel Khuri-Treiman equations and the dashed curve to the
  single-channel solution. The blue dash-dotted curve is the chiral $O(p^4)$
  result.}
\label{fig:eta3piu=t}
\end{figure}
Results for the $\eta\to 3\pi$ amplitude obtained from solving numerically the
Khuri-Treiman equations are presented in fig.~\fig{eta3piu=t} which shows the
real part of the amplitude along the line $t=u$ as a function of $s$. Let us
consider the role of the inelastic channels in four energy regions
\begin{itemize}  
\item[a)] In the neighbourhood of $s=1$ $\hbox{GeV}^2$ there is a very sharp
  energy variation, as one could have easily expected, induced by the
  interference of the $a_0$ and $f_0$ resonances and the presence of the
  $\Kp\Km$ and $\Kz\Kzb$ thresholds.

\item[b)] In the region $0.7\lapprox s \lapprox0.97$ $\hbox{GeV}^2$ the
  coupled-channel amplitude displays a large enhancement as compared to the
  single-channel amplitude.

\item[c)] In the lower energy region, $s \lapprox0.7$ $\hbox{GeV}^2$, on the
  contrary,  the  effect of the inelastic channels is to
  \emph{reduce} the size of the amplitude. One also observes that in this
  region the influence of the inelastic channels becomes small. 
 
\item[d)] In the sub-threshold region, finally, the coupled-channel and
  single-channel amplitudes are essentially indistinguishable. This is
  expected since the amplitudes are constrained to satisfy the same chiral
  matching equations. 
\end{itemize}
\begin{figure}[h]
\centering
\includegraphics[width=0.60\linewidth]{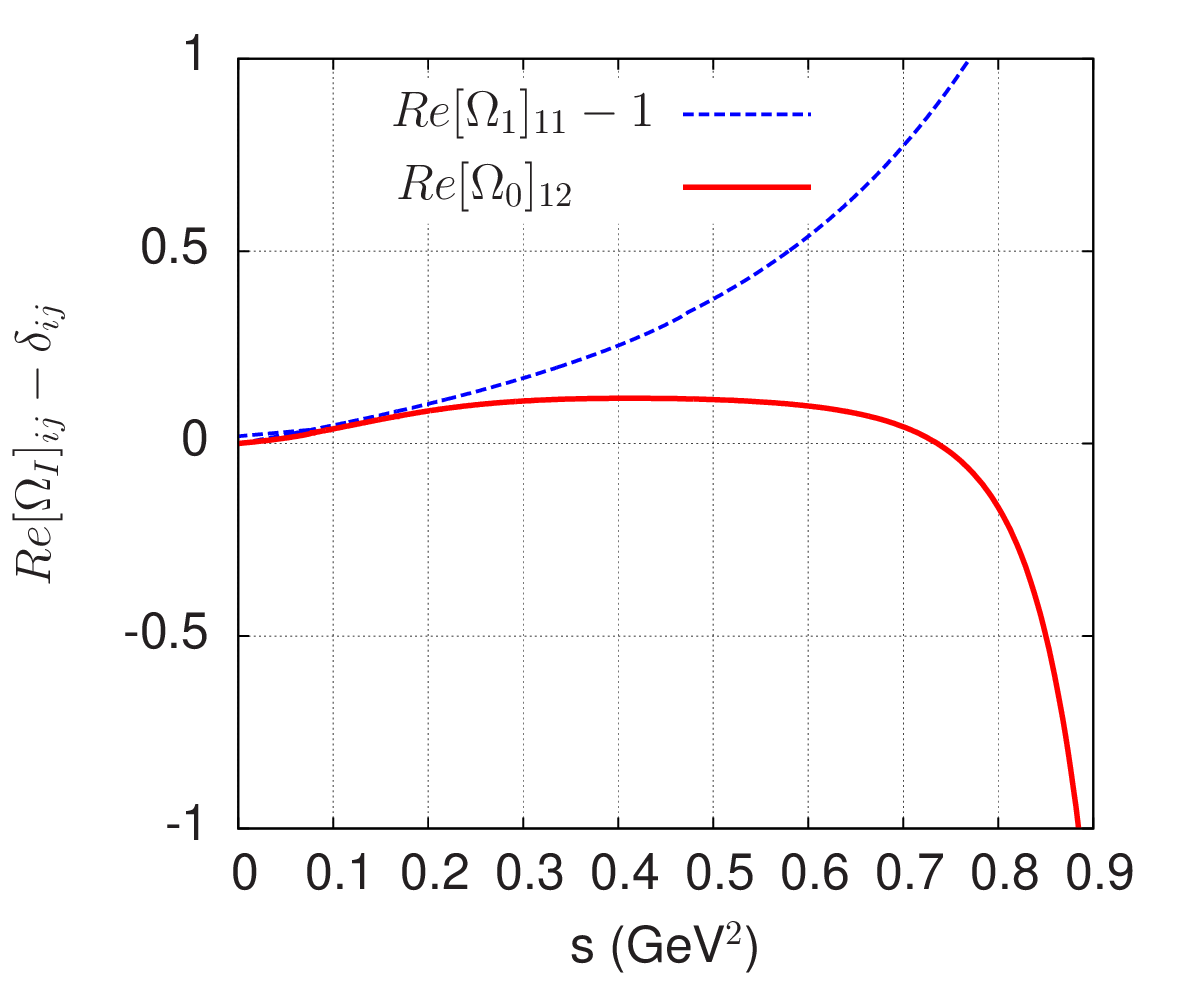}
\caption{\small Components of the Omn\`es matrices $\bm{\Omega}_0$ and
$\bm{\Omega}_1$ which play an important role (see text).}
\label{fig:OM0OM1}
\end{figure}

It is not difficult to identify the main mechanism which generates the 
behaviour described above. For this purpose, let us consider the $i=1,j=1$
component of the matrix $\bm{M_0}$ and let us absorb the the effect of the
integrals $\hat{I}_a$, $\hat{I}_a$ at $s=0$ into the polynomial matrix,
defining 
\be
\tilde{\bm{P}_0}(s)\equiv \bm{P}_0(s)+s^2(\hat{\bm{I}}_a(0)
+\hat{\bm{I}}_b(0))
\en
The real parts of the three components of $\tilde{\bm{P}_0}$ which are
related to the $K\Kbar$ channel have the following expressions
\be
\ba{l}
\re[\tilde{\bm{P}}_0]_{21}(s)\simeq \phantom{-}0.15-4.94\,s-12.3\,s^2\\
\re[\tilde{\bm{P}}_0]_{12}(s)\simeq -0.81+5.73\,s+1.27\,s^2\\
\re[\tilde{\bm{P}}_0]_{22}(s)\simeq
\phantom{-}0.89+0.22\,s-6.25\,s^2\ .\\
\ea
\en
These polynomial coefficients are controlled, we recall, by the matching
conditions to the LO chiral $\eta\pi \to K\Kbar$, $\pi\pi \to K\Kbar$, and 
$K\Kbar\to K\Kbar$ isospin violating amplitudes (see
appendix~\ref{sec:appendixB}). 
The component $\re[\bm{P}_0]_{21}$ is negative in the  region 
$s > 4\mpid$ and dominates the others in the range $s\gapprox 0.2$
$\hbox{GeV}^2$. In fact, the corresponding contribution to
$\bm{M}_0$ dominates in the whole region $4\mpid < s < 0.95$
$\hbox{GeV}^2$. In this region, the contribution from the inelastic
channels  can thus be written approximately as
\be\lbl{M0inel}
[\bm{M}_0]^{inel}_{11}\simeq (\bm{\Omega}_0)_{12} (\bm{\Omega}_1)_{11}
\left(\bm{P}_0 +s^2 (\hat{\bm{I}}_a + \hat{\bm{I}}_b)\right)_{21}
\en
The components of the Omn\`es matrices which appear in eq.~\rf{M0inel}
are plotted in fig.~\fig{OM0OM1}. The salient feature here is that the
real part of the $I=0$ component $(\bm{\Omega}_0)_{12}$ is positive at
low energy and changes sign\footnote{The presence of a zero below the $K\Kbar$
  threshold follows from Watson's theorem which is obeyed by the component
  $(\bm{\Omega}_0)_{12}$ and leads to the relation
  $\re(\bm{\Omega}_0)_{12}/\im(\bm{\Omega}_0)_{12}=\cot\delta_0^0$. This
  implies that $\re(\bm{\Omega}_0)_{12}$ vanishes when the phase shift
  $\delta_0^0$ goes through $\pi/2$.}  at $s\simeq 0.73$ $\hbox{GeV}^2$. This
is the main reason why the inelastic channels decrease the $\eta\to 3\pi$
amplitude below
$0.7$ $\hbox{GeV}^2$ and increase it above. This
behaviour is enhanced by the $I=1$ component $(\bm{\Omega}_1)_{11}$
which is larger than 1 (reflecting the attractive nature of the
$\eta\pi\to \eta\pi$ interaction below 1 GeV).
\begin{figure}[h]
\centering
\includegraphics[width=0.60\linewidth]{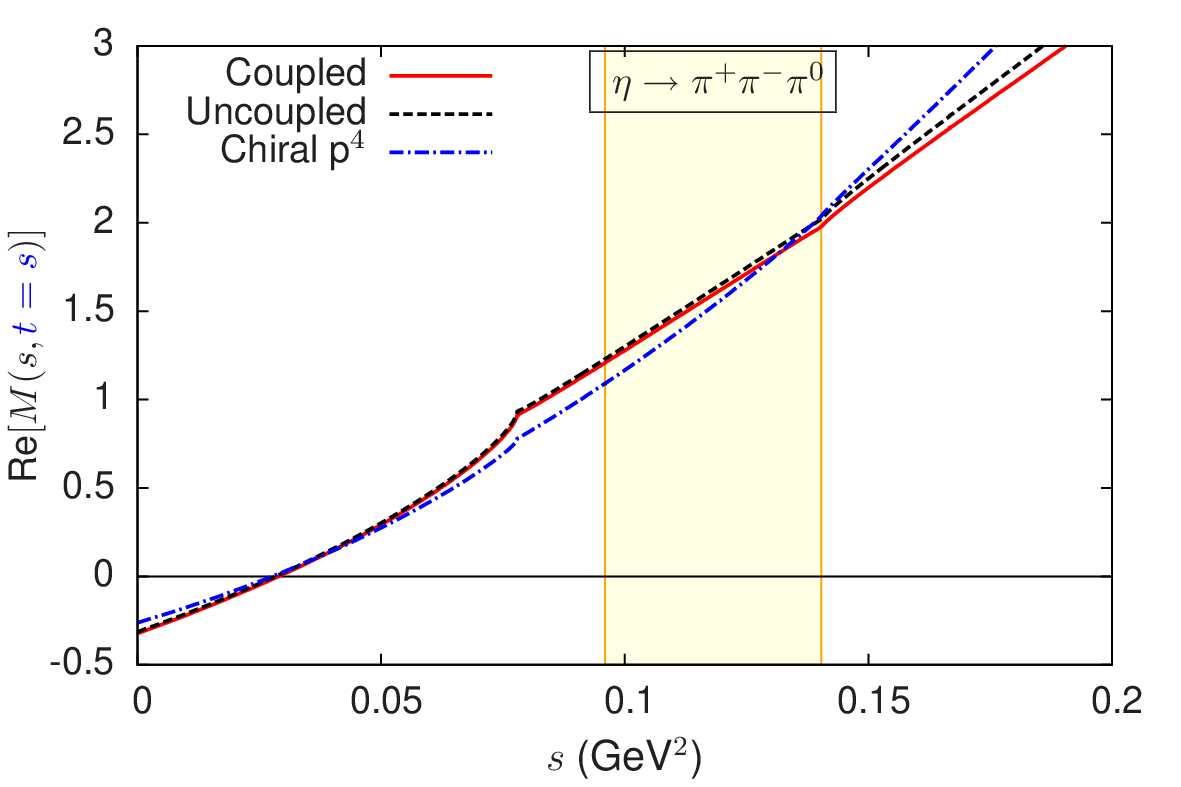}
\caption{\small Real part of the $\eta\to \pi^+\pi^-\pi^0$ amplitude along the
  the $t=s$ line as a function of $s$. The yellow shaded area indicates the
  physical decay region. The lines are as in fig.~\fig{eta3piu=t}.}
\label{fig:eta3piu=s}
\end{figure}

\begin{figure}[h]
\centering
\includegraphics[width=0.60\linewidth]{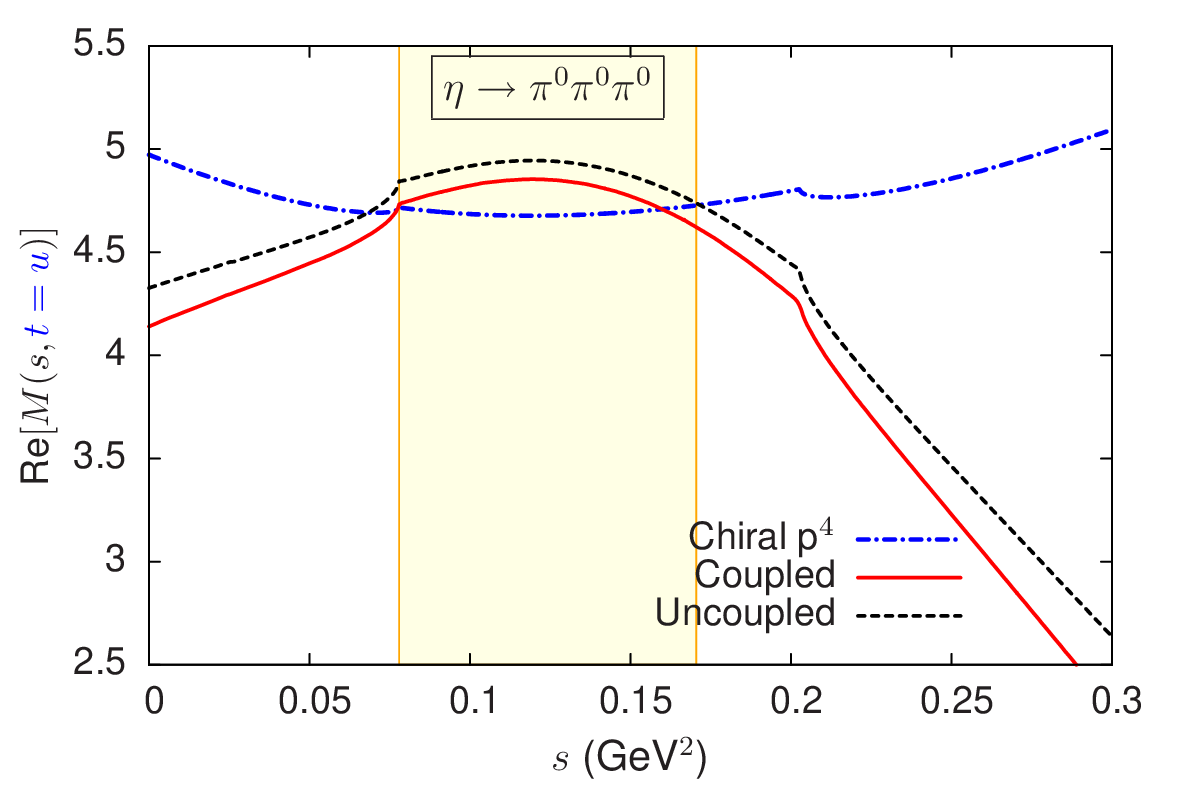}
\caption{\small Real part of the $\eta\to 3\pi^0$ decay amplitude along the
  $t=u$ line as a function of $s$. The lines are as in fig.~\fig{eta3piu=t}.}
\label{fig:eta3pi0}
\end{figure}

The behaviour of the amplitude along the $t=s$ (or $u=s$) line, which displays
the Adler zero, is shown in fig.~\fig{eta3piu=s}. In the sub-threshold region,
the chiral, the single-channel and the coupled-channel amplitudes are seen to
be very close. For the position of the Adler zero, we find
\be
s_A^{NLO}=1.42\,m_\pip^2,\quad
s_A^{SC}= 1.45\,m_\pip^2,\quad
s_A^{CC}= 1.49\,m_\pip^2\ .
\en
Finally, the results for the $\eta\to 3\pi^0$ amplitude are shown in
fig.~\fig{eta3pi0}. The influence of the inelastic channels are seen to be
quite substantial in this case in the whole low-energy region.  One also sees
that there is no region in which there is close agreement between the
dispersive and the chiral $O(p^4)$ amplitudes. This, of course, is because of
the occurrence of the combination $M_0(s)+M_0(t)+M_0(u)$ and the fact that at
least one of the variables $s$, $t$, $u$ must lie above the $\pi\pi$
threshold, thus generating significant $S$-wave rescattering chiral
corrections.

\begin{figure}[ht]
\centering
\includegraphics[width=0.60\linewidth]{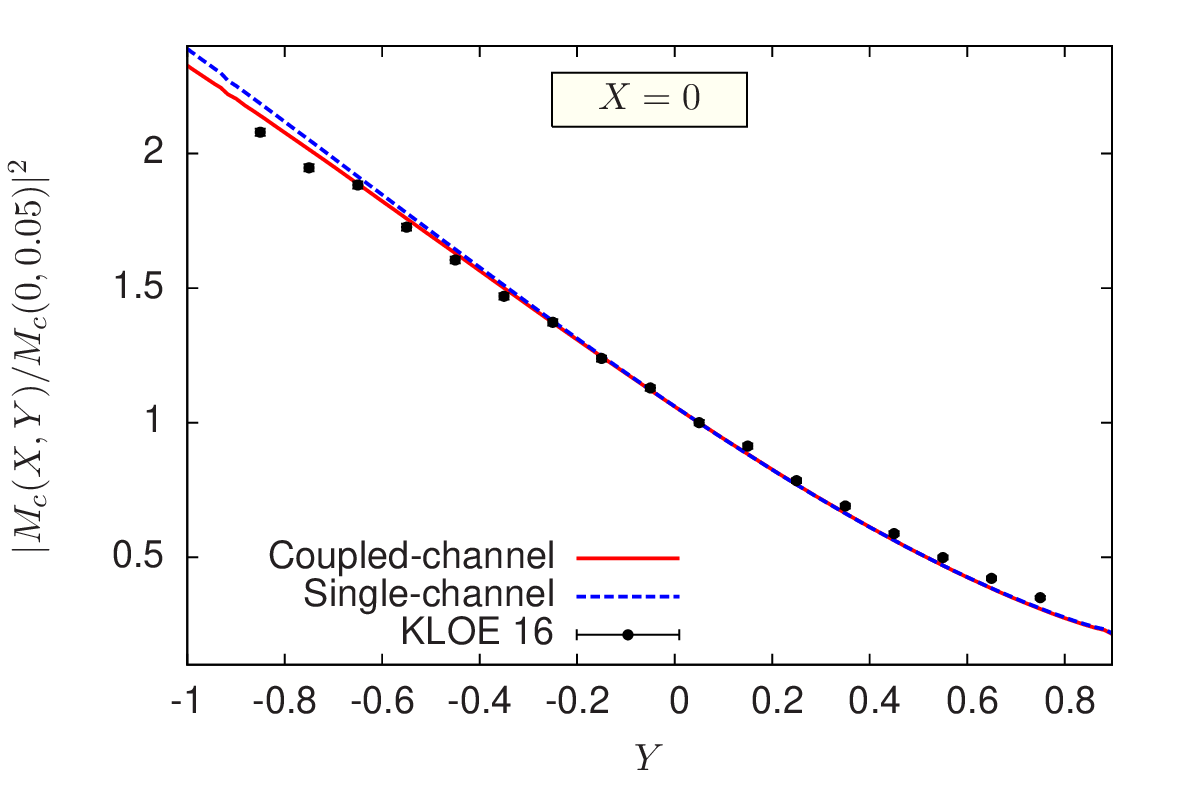}
\includegraphics[width=0.60\linewidth]{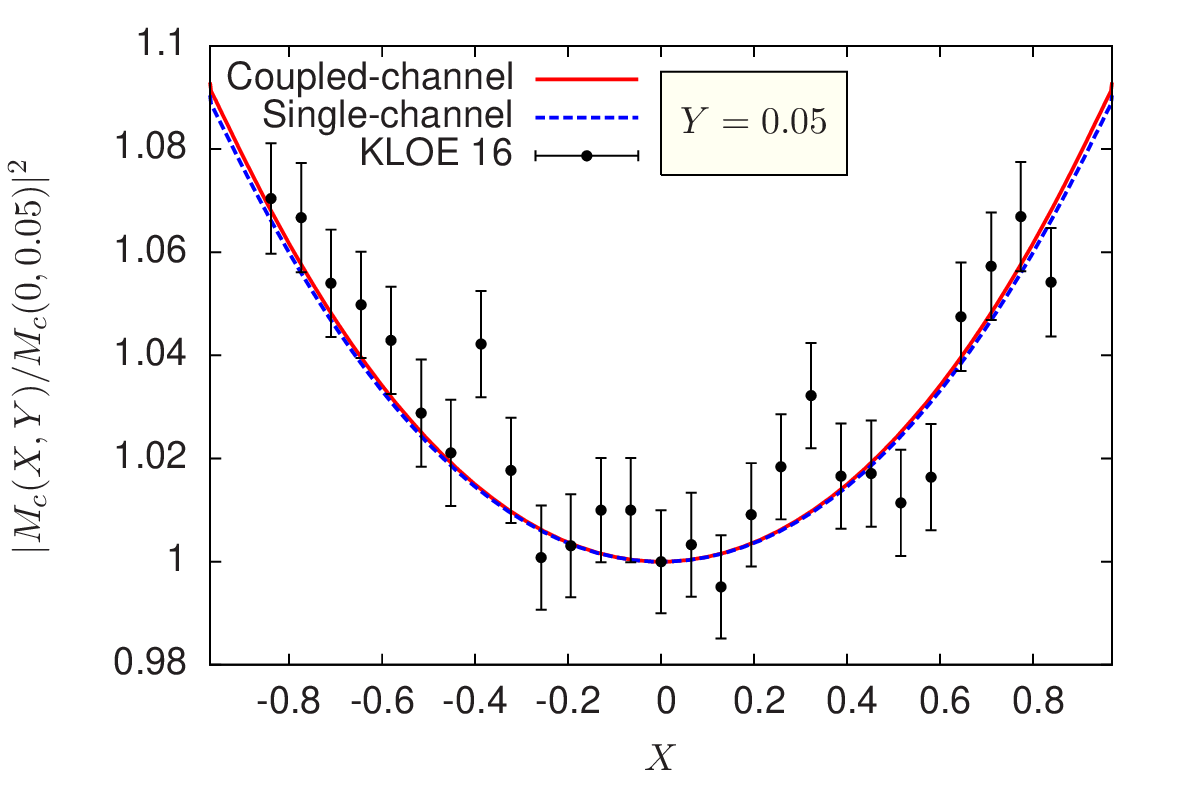}
\caption{\small The computed charged amplitudes squared, normalised to 1 at
  $X=0$, $Y=0.05$ are shown along the line $X=0$ as a function of $Y$ (upper
  plot) and along the line $Y=0.05$ as a function of $X$ (lower plot) and
  compared with the acceptance corrected results provided by the KLOE
  collaboration\cite{Anastasi:2016qvh}.}
\label{fig:dalitzXY}
\end{figure}
\subsection{Dalitz plot parameters}
One traditionally describes the Dalitz plot in terms of two dimensionless
variables $X$, $Y$ such that $|X|,|Y| \le 1$ and the centre of the Dalitz plot
corresponds to $X=Y=0$.  In the case of the charged decay amplitude $\eta \to
\pi^+ \pi^- \pi^0$, the variables $X$, $Y$ are related to the Mandelstam ones
by
\be
X=\frac{\sqrt3}{2\meta\,Q_c}(u-t),\quad
Y=\frac{3}{2\meta\,Q_c}\left( (\meta-m_\piz)^2-s\right)-1
\en
with $Q_c=\meta-2 m_\pip-m_\piz$. Assuming charge conjugation
invariance, the amplitude must be invariant under the transformation $X\to
-X$ and a polynomial parametrisation of the amplitude squared can be
written as
\be\lbl{Mc2XY}
\frac{\vert{M_c}(X,Y)\vert^2}{\vert{M_c}(0,0)\vert^2}
=1+a\,Y +b\,Y^2 +d\,X^2 + f\,Y^3+ g\, X^2 Y +\cdots
\en
In the case of the neutral decay amplitude $\eta\to 3\piz$, $Q_c$ must
be replaced by $Q_n=\meta-3m_\piz$ in the definition of $X$ and
$Y$. Charge conjugation and Bose symmetry imply that the amplitude
must be invariant under the two transformations
\be
z \to z\,\exp\big(\frac{2i\pi}{3}\big),\quad  z\to -z^*
\en
with $z=X+iY$. The amplitude squared can thus be represented as 
\be\lbl{Mn2XY}
\frac{\vert{M_n}(X,Y)\vert^2}{\vert{M_n(0,0)}\vert^2}=1+2\alpha\, |z|^2 
+2\beta\, \im(z^3)+\cdots
\en

A direct comparison of the dispersive amplitudes squared with the experimental
data from KLOE~\cite{Anastasi:2016qvh} is shown in fig.~\fig{dalitzXY} and
our numerical results for the Dalitz plot parameters are collected in
table~\Table{Dalitzpar}.  The numbers quoted in the table are obtained in a
way which parallels the experimental determination: a discrete binning of the
Dalitz plot is performed (with a few hundred bins) and a global least squares
fit of the theoretical (chiral or dispersive) amplitudes squared is performed
using the representations~\rf{Mc2XY},~\rf{Mn2XY}.  The table also shows the
two most recent experimental
determinations~\cite{Ablikim:2015cmz,Anastasi:2016qvh}.
\begin{table}[h]
\centering
\bt{c||r|r|r||r|r}\hline\hline
\TT\BB$\eta\to \pi^+\pi^-\pi^0$ & $O(p^4)$ & single-ch. & coupled-ch. & KLOE
& BESIII\\ \hline
\TT a &-1.328 & -1.156 & -1.142 & -1.095(4)   & -1.128(15) \\ 
    b & 0.429 &  0.200 &  0.172 &  0.145(6)   &  0.153(17) \\  
    d & 0.090 &  0.095 &  0.097 &  0.081(7)   &  0.085(16)\\  
    f & 0.017 &  0.109 &  0.122 &  0.141(10)  &  0.173(28)\\  
\BB g &-0.081 & -0.088 & -0.089 & -0.044(16)  & 
 --$\phantom{3(28)}$\\  
\hline
\TT\BB$\eta\to\pi^0\pi^0\pi^0$ &           &            &
&\multispan{2}\hfil PDG \hfil \\ \hline
\TT
$\alpha$ & +0.0142 & -0.0268 & -0.0319   & \multispan{2}\hfil-0.0318(15)\hfil\\
$\beta$  & -0.0007 & -0.0046 & -0.0056   & \multispan{2}\hfil$-$\hfil\\
 \hline\hline
\et
\caption{\small Results for the  Dalitz plot parameters
of the charged and neutral $\eta\to 3\pi$ decays
  based on the NLO chiral amplitude (in the form given in
  appendix~\ref{sec:appendixA}) 
  and its dispersive extrapolations based on single-channel and
  coupled-channel Khuri-Treiman equations and a matching procedure
  described in the text. The last two columns show experimental
  results from refs.~\cite{Anastasi:2016qvh,Ablikim:2015cmz}
  and~\cite{Olive:2016xmw}. } 
\lbltab{Dalitzpar}
\end{table}

It is clear, at first, that the amplitudes obtained from solving the
Khuri-Treiman equations and constrained to match the chiral NLO
amplitudes are in much better agreement with the experimental results
in the physical decay region than the NLO amplitude itself. In
particular, the parameter $b$ which was too large by a factor of three
is reduced by a factor of two and the parameter $\alpha$ which was
positive becomes negative. This is in close agreement with the results
obtained a long time ago by Kambor
{\it et al.}~\cite{Kambor:1995yc}. 
Our main new result is that taking into account the $K\Kbar$ inelastic
channels and the effect of $\eta\pi$ rescattering has a non negligible
influence on the Dalitz parameters and tends to further improve the agreement
with experiment. The influence of these inelastic channels for the parameters
$d$ and $g$ is small (less than 5\%) but quite significant for the parameter
$b$ which is reduced by 17\% and now lies within 15\% of the experimental
value. This reflects the reduction of the amplitude caused by the inelastic
effects at low energy discussed in sec.~\ref{sec:inelrole}.
Similarly, the parameter
$\alpha$ is modified by approximately 20\% by the $K\Kbar$ inelastic channels
and becomes rather close  to the experimental result. The
parameter $g$ is the only one which shows a mismatch, by a factor of two, with
the value measured by KLOE.

Finally, let us consider the sensitivity of the Dalitz plot parameters to the
strength of the $\eta\pi$ interaction, which is not precisely known at
present. This is illustrated in table~\Table{etapirole}. The table also shows
that varying the $O(p^4)$ coupling $L_3$ has a significant influence, in
particular on the parameter $d$.
\begin{table}[h]
\centering
\bt{cccc}\hline\hline
\TT\BB      & no $\eta\pi$ & large $\eta\pi$   & $L_3=-2.65\cdot10^{-3}$\\ \hline
\BB $\Delta{a}/\vert{a}\vert\ $     & $-0.6\%$ & $+0.8 \%$ & $+3.9 \%$   \\
\BB $\Delta{b}/\vert{b}\vert\ $     & $+9.0\%$ & $-9.6 \%$ & $-2.4 \%$   \\
\BB $\Delta{d}/\vert{d}\vert\ $     & $-0.7\%$ & $+0.8 \%$ & $-13.0\%$   \\
\BB $\Delta{f}/\vert{f}\vert\ $     & $-6.3\%$ & $+6.4 \%$ & $-11.3\%$   \\
\BB $\Delta{g}/\vert{g}\vert\ $     & $-0.2\%$ & $+0.3 \%$ & $+10.8\%$   \\
$\Delta{\alpha}/\vert{\alpha}\vert$ & $+9.1\%$ & $-9.2\%$  & $+5.5 \%$   \\ \hline\hline
\et
\caption{\small Relative variation of the Dalitz plot parameters from their
  central values. Second column: the $\eta\pi$ $T$-matrix elements are set to
  zero, third column: the $\eta\pi$ $T$-matrix has larger phase-shifts:
  $\delta_{\eta\pi\to K\Kbar}(m_{a_0(1450)})=180\degree$ (instead of
  $100\degree$) and the scattering length is $a_0^{\eta\pi}=21.6\cdot10^{-3}$
  (instead of $13.2\cdot10^{-3})$), fourth column $L_3=-2.65\cdot10^{-3}$
  (from ref.~\cite{Colangelo:2015kha}) instead of $L_3=-3.82\cdot10^{-3}$. }
\lbltab{etapirole}
\end{table}

\subsection{The ratio $\Gamma(\eta\to 3\pi^0)/\Gamma(\eta\to \pip\pim\piz)$}
Let us quote here the results for the ratio of the $3\piz$ and the
$\pip\pim\piz$ decay rates 
\be
{\cal R}_{3\piz/\pip\pim\piz}\equiv \frac{\Gamma(\eta\to
  3\piz)}{\Gamma(\eta\to\pip\pim\piz)} 
\en
We find that the influence of the inelastic
channels on this ratio is very small, 
\begin{align}
{\cal R}_{3\piz/\pip\pim\piz}\simeq &
1.451\,(\hbox{coupled-channel})\ ,\nonumber\\
{\cal R}_{3\piz/\pip\pim\piz}\simeq & 1.449\,(\hbox{single-channel})
\end{align}
As compared with the chiral
$O(p^4)$ result  
\be
{\cal R}_{3\piz/\pip\pim\piz} =1.425\,(\hbox{Chiral
  NLO})\ ,
\en
this ratio is thus predicted to increase under the effect of the
final-state interactions, by only 2\%. The experimental status of
this quantity is not completely clear at present, as the
PDG\cite{Olive:2016xmw} quotes two different numbers
\be
{\cal R}_{3\piz/\pip\pim\piz}=
1.426\pm0.026\,(\hbox{PDG fit})\ ,\quad
1.48\pm 0.05\,(\hbox{PDG average})\ .
\en
Besides, the CLEO collaboration~\cite{Lopez:2007ab} has performed an
experiment dedicated to the determinations of the  $\eta$ meson decay
branching fractions and they quote 
\be
{\cal R}_{3\piz/\pip\pim\piz}= 1.496\pm 0.043 \pm 0.035\ (\hbox{CLEO})
\en
as the most precise determination of the $3\piz/\pip\pim\piz$ ratio.

\subsection{The quark mass ratio $Q$ from the chirally matched dispersive
  amplitude}\label{sec:Qremark} 
It must first be
reminded that, once the electromagnetic interaction is taken into account, the
quark mass ratio $Q$ is no longer invariant under the QCD renormalisation group
since the quark mass variation with the scale depends on its electric charge
\be
\mu_0\frac{ dm_q(\mu_0)}{d\mu_0} =-\left(\gamma^{QCD} +
e^2_q\,\gamma^{QED}\right)\, m_q(\mu_0)\ ,\quad
\gamma^{QED}= \frac{3 e^2}{8\pi^2} + O(e^4)\ 
\en
(with $e_u=2/3$, $e_d=e_s=-1/3$).
This implies the following scale variation for the factor 
$\epsilon_L$ 
\be
\mu_0\frac{ d\epsilon_L}{d\mu_0}=
\frac{e^2}{16\pi^2}\frac{\mpid}{3\sqrt3\fpid} + O(e^2(m_d-m_u), e^2
m_q^2 )\ .
\en
It can then easily be verified, using the equations of
appendix~\ref{sec:appendixA} 
which include the $e^2\mpid$ contributions~\cite{Baur:1995gc}, that
the scale invariance of the complete NLO chiral amplitude is restored
thanks to the combination of two of the electromagnetic coupling
constants~\cite{Urech:1994hd}, $K_9^r+K_{10}^r$. Indeed, as shown in
refs.~\cite{Bijnens:1996kk,Moussallam:1997xx} this combination depends
not only on the  chiral scale $\mu$ but also on the QCD scale
$\mu_0$ and satisfies
\be
\mu_0\frac{d}{d\mu_0}\left(
K_9^r(\mu,\mu_0)+K_{10}^r(\mu,\mu_0)\right)=\frac{3}{4}\frac{1}{16\pi^2}\ .
\en
In practice, this means that in order to determine $Q(\mu_0)$ from the
$\eta\to 3\pi$ amplitude we must specify the values of the electromagnetic
chiral couplings $K_i$ at the corresponding scale.  We choose here
$\mu_0=0.77$ GeV and estimate the values of the couplings $K_i^r(\mu,\mu_0)$
from a resonance saturation model~\cite{Ananthanarayan:2004qk}.  Such
estimates are qualitative at best but it can be shown, based on general order
of magnitude arguments, that the uncertainty induced on the amplitude should
not exceed a few percent~\cite{Baur:1995gc,Ditsche:2008cq}.

Having verified that the dispersive amplitude is in qualitative agreement with
experiment concerning the Dalitz plot parameters, we can make an estimate of
the value of $Q$.  In the present approach, the amplitude in the physical
decay region is derived as a Khuri-Treiman solution uniquely defined from the
chiral NLO amplitude by the set of
four matching equations, and thus has a definite dependence on $Q$.  We can
then estimate the value of $Q$ by the requirement that the dispersive
amplitude reproduces the experimental values of the $\eta\to 3\pi$ decay
widths
\be\begin{array}{ll}
\Gamma_{exp}[\,\eta\to \pip\pim\piz]=& (299\pm 11)\ \hbox{eV} \\
\Gamma_{exp}[\,\eta\to 3\piz]       =& (427\pm 15)\ \hbox{eV} \\
\end{array}\en
taken from the PDG~\cite{Olive:2016xmw} (constrained fit).
Doing this, we find
\be\lbl{Q2results}
\begin{array}{lll}
\eta\to \pip\pim\piz: & Q=21.8\pm 0.2\hbox{ (single-channel), }
                        Q=21.6\pm 0.2\hbox{ (coupled-channel)} \\
\eta\to 3\pi^0 :       & Q=21.9\pm0.2\hbox{ (single-channel), }
                        Q=21.7\pm 0.2\hbox{ (coupled-channel)}\\
\end{array}
\en
where the quoted errors only reflect the experimental ones on the widths (we
comment below on the theoretical error).  This shows
that the effect of the inelastic channels on the determination of $Q$ is
rather small, of the order of 1\%, and tends to decrease its value. 

The central value of $Q$ is somewhat smaller than the results which are
obtained from lattice QCD+QED simulations of hadron masses:
$Q=22.9\pm0.4$ (ref.~\cite{Horsley:2015vla}), $Q=23.4(0.4)(0.3)(0.4)$
(ref.~\cite{Fodor:2016bgu}). The error on $Q$ associated with the phase-shifts
below 1 GeV is rather small, of the order of 1\%. The error associated with
the NLO amplitude, essentially related to $L_3$, is of the same order.  The
largest error arises from chiral NNLO contributions to the amplitude which
will modify the determination of the polynomial parameters via the matching
conditions. Assuming that they induce a 10\% relative error in the
determination of each one of the four polynomial parameters $\alpha_0$,
$\beta_0$, $\gamma_0$, $\beta_1$ and assuming the errors to be independent,
gives the following theory error on $Q$
\be\lbl{errorQ} 
\Delta{Q}_{th}=\pm 2.2\   
\en 
$Q$ being mostly sensitive to the variation of the first two parameters
$\alpha_0$, $\beta_0$. We have also varied the remaining 14 polynomial
parameters, assuming a 20\% relative error, and found that they have a much
smaller influence.  
Within the error~\rf{errorQ}, the determination based on $\eta\to 3\pi$ decay
is compatible with the lattice QCD results, which confirms that the size of
the NNLO corrections to the $\eta$ decay amplitude in the sub-threshold region
should not exceed $10\%$.

\subsection{Further experimental constraints on $Q$}
Our estimate for the error on $Q$ (eq.~\rf{errorQ}) was based on a 
general order of magnitude assumption on the size of the NNLO corrections to
the four leading polynomial parameters.
More precise information on the size of these corrections can be derived by
making use of the precise experimental results on the energy
dependence across the Dalitz region, imposing that the dispersive amplitude
reproduces these via a least-squares fit.
Not all the four leading polynomial parameters can get independently
constrained in this way since a ratio of amplitudes is involved. We will make
the simple choice to fix one of them, $\alpha_0$, from its chiral value and to
perform a variation of the three others $\beta_0$, $\gamma_0$, $\beta_1$.
We will use the latest KLOE experimental data~\cite{Anastasi:2016qvh} which
consist of a set of amplitudes squared, $\vert{\cal T}_{exp}(X_i,Y_i)\vert^2$,
measured over 371 energy bins in the Dalitz region and satisfying the
normalisation condition
\be
\vert{\cal
  T}_{exp}(0,0.05)\vert^2=1\pm0.01\ .
\en 
We introduce corresponding theory
amplitudes ${\cal T}_{th}(X,Y)= M_c(X,Y)/M_c(0,0.05)$ and define the
$\chi^2$ as
\be\lbl{chi2def}
\chi^2=\sum_{bins}\frac{ \left(
\lambda \vert {\cal T}_{th}(X_i,Y_i)\vert^2 
- \vert{\cal T}_{exp}(X_i,Y_i)\vert^2
\right)^2}
{\left\vert \Delta{\cal T}_{exp}(X_i,Y_i)  \right\vert^2}
\en
allowing for a floating of the normalisation within the experimental error via
a parameter $\lambda$. 

At first, setting $\lambda=1$ and computing the $\chi^2$ with our chirally
matched central amplitude with $L_3=-3.82\cdot10^{-3}$ we obtain
$\chi^2=3079$. If, instead, we use the value recently derived in
ref.~\cite{Colangelo:2015kha} from $K_{l4}$ decays: $L_3=-2.65\cdot10^{-3}$,
one obtains a significantly reduced result: $\chi^2=714$. It thus seems
reasonable to use this value as a starting point, which fixes the central
value of $\alpha_0$.  We have then searched for a minimum of the $\chi^2$ by
varying the real parts of the three polynomial parameters $\beta_0$,
$\gamma_0$, $\beta_1$ (keeping their imaginary parts fixed) and the
normalisation $\lambda$. The other polynomial parameters, in particular
$\alpha_0$, are kept fixed to their chirally-matched value. In this way, we
obtain at the minimum: $\chi^2_{min}=387$, which corresponds to a value per
degree of freedom $\chi^2_{min}/dof= 1.055$.  The fitted values of the
polynomial parameters differ from the chirally-matched values by less than 5\%
(the numerical values are given in appendix~\ref{sec:supplement}). This
confirms that this difference can be consistently attributed to chiral NNLO
effects.

The theory error in this approach is dominated by the variation (by 10\%) of
the single polynomial parameter $\alpha_0$ (we have also added an error
associated with the input $T$-matrices and an error from varying the imaginary
parts of the polynomial parameters). The error induced by the variation of the
parameters $\beta_0$, $\gamma_0$, $\beta_1$ is now computed using the
covariance matrix as evaluated by the MINUIT fitting
program~\cite{James:1975dr} (this matrix is given in
appendix~\ref{sec:supplement}). This error is added in quadrature with the
experimental error on the $\pi^+\pi^-\pi^0$ and $3\pi^0$ decay rates.
Finally, the result for the quark mass ratio as determined from this fitted
amplitude can be written as
\be\lbl{Qfit}
Q_{fit}= 21.50 \pm 0.70_{th} \pm 0.67_{exp}\ .
\en
We find it useful to quote the theoretical and experimental errors separately
since the former is not necessarily gaussian. 
Previous determinations of $Q$ which combine chiral constraints and fits to
Dalitz plot data have been performed in
refs.~\cite{Martemyanov:2005bt,Kampf:2011wr,Colangelo:2011zz,Guo:2015zqa,Guo:2016wsi,Colangelo:2016jmc}. 
In the most sophisticated of these approaches~\cite{Colangelo:2016jmc}, five
polynomial parameters are included in the fit and the effect of the
$\pip-\piz$ mass difference in the amplitudes is accounted for.

Finally, we quote the values of the $3\pi^0$ Dalitz plot parameters which can
be predicted from our fitted amplitude
\be
\alpha_{fit}=-0.0337\pm 0.0012,\quad
\beta_{fit} =-0.0054\pm 0.0001\ .
\en

\section{Conclusions}
We have proposed an extension of the Khuri-Treiman formalism for the
$\eta\to 3\pi$ amplitude
which includes the $\eta\pi$ and the $K\Kbar$ channels in the unitarity
equations in addition to the elastic $\pi\pi$ channel.
Modulo some approximations (in particular we do not attempt to impose
unitarity in the crossed channels involving kaons like $\pi K\to \pi K$ or
$\eta K \to \pi K$) the equations for the one-variable functions $M_0$ and
$M_2$ are shown to be simply replaced by $2\times2$ matrix equations. These
are given in eqs.~\rf{M0disp} and~\rf{M2disp} which involve both the $I=0$ and
the $I=1$ Omn\`es $2\times2$ matrices. Eq.~\rf{M0disp} exhibits explicitly the
contribution induced by the physical $K^0-K^+$ mass difference via
unitarity in integral form.

The amplitudes derived from this extended framework should be valid in an
energy range which covers the physical decay region and also the physical
region of the scattering $\eta\pi\to \pi\pi$ below 1 GeV.  Given a fixed
number of polynomial parameters, an improved precision at low energy should
result from the fact that the effects of the two prominent light scalar
resonances $a_0(980)$ and $f_0(980)$ are taken into account in the dispersive
integrals.

Using four polynomial parameters in the $\eta\to 3\pi$ amplitude we have 
reconsidered the idea of performing a  prediction of the amplitude in
the physical region as an extrapolation of the  $O(p^4)$ chiral amplitude,
uniquely defined by fixing all the polynomial parameters by  matching
conditions. 
These are imposed in the form of a set of equations which insure that the
differences between the dispersive and the $O(p^4)$ chiral $\eta\to 3\pi$
amplitude are of order $p^6$ or higher. One verifies then, that the chiral and
the dispersive $\eta\to \pi^+\pi^-\pi^0$ amplitudes are very close in the
neighbourhood of the Adler zero. These conditions also insure, for the charged
decay amplitude, that the single and multi-channel dispersive amplitudes are
quasi-identical in the whole region $0\le s\le 4\mpid$, $|t-u| \le
(\meta+\mpi)^2$. In contrast, for $\eta\to 3\pi^0$, one finds that the
unitarity induced chiral corrections are significant even in the sub-threshold
region.

We have considered the Dalitz plot parameters and we found that the induced
influence of the $a_0(980)$, $f_0(980)$ resonances is not negligible, in
particular for the neutral mode. The modifications of the parameters, in the
coupled-channel framework, go in the sense of improving the agreement with
experiment, in particular for the parameters $a$, $b$, $f$ of the charged
mode. The parameter $\alpha$, for the neutral mode is modified by 20\% by the
effects of the resonances and lies rather close to the experimental value. The
remaining differences between the experimental and the dispersive-theoretical
amplitude suggest that NNLO contributions are needed in the matching
conditions, at the 5-10\% level, which seems quite plausible. Some of these
NNLO effects could be accounted for in a more general framework which would
implement both unitarity and crossing symmetry completely for all the channels
involved. This is left for future work.

The $\eta\to 3\pi$ amplitudes constructed in the present approach inherit a well
defined dependence on the quark mass ratio $Q$ from that of the chiral NLO
amplitude. We can then determine $Q$ such as to reproduce the integrated decay
widths.  The central value that one obtains is somewhat low compared to the
recent determinations from lattice QCD simulations but it is compatible within
the uncertainty induced by the NNLO effects in the matching. Some knowledge of
these NNLO effects seems necessary in order to improve the precision of the
determination of $Q$ by this method.

\section*{Acknowledgements}
B.~M. acknowledges rewarding discussions with \'Emilie Passemar.
M.~A.\ acknowledges financial support from the Spanish MINECO and
European FEDER funds under the contracts 27-13-463B-731 (``Juan de la
Cierva'' program), FIS2014-51948-C2-1-P, FIS2014-57026-REDT, and
SEV-2014-0398, and from Generalitat Valenciana under contract
PROMETEOII/2014/0068.
\appendix
\section{The amplitude $\eta\to \pip\pim\piz$  at chiral order $p^4$ and
  $e^2p^2$}\label{sec:appendixA} 
We give below the explicit expressions for the three functions $\bar{M}_I$ in
the chiral expansion at order $p^4$ and including $e^2 p^2$ contributions
as used in the present work. They are given in a form which is manifestly
scale invariant independently of the values of $m_\pi$, $m_K$, $m_\eta$. The
amplitude is identical with the one originally computed in
ref.~\cite{Gasser:1984pr} when $e^2=0$ and  the $m_\pi$, $m_K$, $m_\eta$
values in the NLO part obey the Gell-Mann-Okubo relation
$3\metad=4\mkd-\mpid$. Following
ref.~\cite{Gasser:1984pr} the LEC's $L^r_5$, $L^r_7$, $L^r_8$ are expressed
in terms of the two physical quantities\footnote{In eq.~\rf{DeltaGMO}
  $\bar{m}_P$ are QCD masses for which we use the values provided by the FLAG
  review~\cite{Aoki:2013ldr} $\bar{m}_\pi=0.1348$, $\bar{m}_K=0.4942$ and
  $\bar{m}_\eta\simeq m_\eta=0.537862$ (all in GeV) which gives
  $\Delta_{GMO}=0.2068$. Also using $F_K/F_\pi$ from this review gives
  $\Delta_F=0.2005$. Elsewhere in the chiral formulae we use $F_\pi=92.21$ MeV
  (from the PDG), $m_\pi=0.13957$, $m_K=m_\Kp=0.493677$ GeV
  and $L_3=-3.82\cdot10^{-3}$.  }
\be\lbl{DeltaGMO}
\Delta_{GMO}=\frac{4\bar{m}^2_K-\bar{m}^2_\pi-3\bar{m}^2_\eta}
{\bar{m}^2_\eta- \bar{m}^2_\pi},\quad
\Delta_F=\frac{F_K}{F_\pi}-1
\en
and in terms of the  quark mass ratio $(m_s+\hat{m})/\hat{m}$ (see
ref.~\cite{gl85}). We also include the electromagnetic contributions
of order $e^2 p^2$ evaluated in ref.~\cite{Baur:1995gc} which allows
one to identify the piece induced by the physical $K^+-K^0$ mass
difference via unitarity. Further electromagnetic corrections
which have been computed in ref.~\cite{Ditsche:2008cq} are not included here.
The expressions for $\bar{M}_I$ given below also
implement the $w=0$ conditions~\rf{MIzero}, which simplifies somewhat the
writing of the matching relations. 

Using the following notation
\be
\Delta_{PQ}= m^2_P-m^2_Q,\quad
R_{PQ}= \frac{ m_P^2  }{\Delta_{PQ}}\,\log\frac{m^2_P}{m^2_Q}
\en
the function $\bar{M}_0$ reads:
\begin{align}\lbl{M0bar}
\bar{M}_0(s) = &
       + {(3 \,s-4 \,m_\pi^2)\over \Delta_{\eta\pi}}  \,\Big(
            1
          + \frac{2}{3} \,{\Delta_{GMO}}
          + 2  m_\pi^2 \,{\Delta_{F}\over \Delta_{K\pi}}
       \Big)+ {m_\pi^2\over \Delta_{\eta\pi}}  \,\Big(
          - \frac{8}{3} \,{\Delta_{GMO}}
       \Big)\nonumber\\ &
       + \frac{2}{3}\,{m_\pi^2 \,m_K^2\over \Delta_{\eta\pi} \,\fpid}  \,
            \bar{J}'_{\pi\eta}(0)\,\Big(
          \,{m_{\eta}^2+3\,m_{\pi}^2-\frac{5}{3} \,s} \Big)\nonumber\\ &
       + {1\over16\pi^2} \,{m_\pi^2 \,m_K^2\over \Delta_{\eta\pi}
         \,\fpid}  \,\Big( 
          - 4
          + 2 \,R_{\pi\eta}
          + \frac{4}{3} \,R_{\pi K}
          - \frac{8}{3} \,\log({m_\pi^2\over m_{\eta}^2})
          + 24 \,\log({m_{\eta}^2\over m_K^2})
       \Big)\nonumber\\ &
       + {1\over16\pi^2} \,{m_\pi^4\over \Delta_{\eta\pi} \,\fpid}  \,\Big(
          + 4
          + 16 \,R_{\pi\eta}
          - \frac{58}{3} \,R_{\pi K}
          + 2 \,\log({m_\pi^2\over m_K^2})
          + \frac{28}{3} \,\log({m_{\eta}^2\over m_K^2})
       \Big)\nonumber\\ &
       + {1\over16\pi^2} \,{s \,m_K^2\over \Delta_{\eta\pi} \,\fpid}  \,\Big(
          + 4
          - 2 \,R_{\pi\eta}
          - 12 \,\log({m_{\eta}^2\over m_K^2})
       \Big)\nonumber\\ &
       + {1\over16\pi^2} \,{s \,m_\pi^2\over \Delta_{\eta\pi} \,\fpid}  \,\Big(
          - 4
          - \frac{11}{2} \,R_{\pi\eta}
          + \frac{15}{2} \,R_{\pi K}
          + 7 \,\log({m_\pi^2\over m_{\eta}^2})
          + 4 \,\log({m_{\eta}^2\over m_K^2})
       \Big)\nonumber\\ &
       + {1\over16\pi^2} \,{s^2\over \Delta_{\eta\pi} \,\fpid}  \,\Big(
          - \frac{3}{4}
          + \frac{3}{4} \,{m_\pi^2\over m_K^2}
          - 3 \,\log({m_\pi^2\over m_K^2})
       \Big)\nonumber\\ &
       + {m_\pi^2\over \Delta_{\eta\pi} \,\fpid}  \,\Big(
          + \frac{2}{9} \,\bar{J}_{\pi\eta}(s) \,(2 \,m_K^2 - 6
          \,m_\pi^2 + 3 \,s) 
          - \frac{2}{3} \,\bar{J}_{\eta\eta}(s) \,(m_K^2 - m_\pi^2)
       \Big)\nonumber\\ &
       + {1\over \Delta_{\eta\pi} \,\fpid}  \,\Big(
          - \frac{1}{3} \,\bar{J}_{\pi\pi}(s) \,(m_\pi^2 - 2 \,s) \,(2
          \,m_K^2 - 6 \,m_\pi^2 + 3 \,s) \nonumber\\ &
         \quad\qquad - \frac{1}{6} \,\bar{J}_{KK}(s) \,(8 \,m_K^4 - 12 \,s
          \,m_K^2 - 6 \,s \,m_\pi^2 + 9 \,s^2) 
       \Big)\nonumber\\ & 
       +\Delta\bar{M}_0^a(s)+\Delta\bar{M}_0^b(s)+\Delta\bar{M}_0^c(s)\ .
\end{align}
The contributions $\Delta\bar{M}_0^a$, $\Delta\bar{M}_0^b$ are induced by the
electromagnetic interaction and proportional to $e^2$,
\begin{align}\lbl{DeltaM0ab}
\Delta\bar{M}_0^a(s)= & -\frac{1}{\epsilon_L}\,\frac{4\,e^2\,\mpid}
{9\sqrt3\fpid}\,\frac{(3\,s-4\,\mpid)}{\Delta_{\eta\pi}}\Big(
-\frac{3}{2}\,(-2K^r_3+K^r_4)-K^r_5-K^r_6\nonumber\\ 
\ & +K^r_9+K^r_{10} +\frac{3\,C}{2\fpiq}\frac{1}{16\pi^2}(1+L_K)
\Big)\nonumber\\ 
\Delta\bar{M}_0^b(s)= & \frac{1}{\epsilon_L}\,
\frac{e^2\,C }{3\sqrt3\,\fpis}\Big((4\,\mkd - 3\,s)\,\bar{J}_{KK}(s)
+3\mkd\, \bar{J}'_{KK}(0)(\frac{5}{3}\,s-\metad-3\mpid)
\Big)\ .
\end{align}
The last contribution, $\Delta\bar{M}_0^c$, is induced by the physical
$K^+-K^0$ mass difference
\be\lbl{DeltaM0c}
\Delta\bar{M}_0^c(s)=-\frac{1}{\epsilon_L}\frac{1}{16\sqrt3 \fpiq}\,
s(4\,\mkd-3\,s)\left(\bar{J}_{K^0K^0}(s)- \bar{J}_{K^+K^-}(s)\right)\ .
\en
The parameters $C$ and $K^r_i$ in the above expressions are the chiral
coupling constants which appear at order $e^2$ and $e^2\, p^2$
respectively~\cite{Urech:1994hd}. 

The chiral expression for the function $\bar{M}_1$ reads
\begin{align}\lbl{M1bar}
\bar{M}_1(t) = &
       + t \,{1\over \Delta_{\eta\pi} \,\fpid}  \,\Big(
          - 4 \,L^r_{3}+{1\over16\pi^2} \Big( \frac{1}{4}
          - \log({m_\pi^2\over m_K^2})
       \Big)
       \Big)\nonumber\\ &
       + {1\over \Delta_{\eta\pi} \,\fpid}  \,\Big(
          - \frac{1}{4} \,\bar{J}_{\pi\pi}(t) \,(4 \,m_\pi^2 - t)
          - \frac{1}{8} \,\bar{J}_{KK}(t) \,(4 \,m_K^2 - t)\ 
       \Big) 
\end{align}
and it has no electromagnetic contributions. The function $\bar{M}_2$,
finally, reads
\begin{align}\lbl{M2bar}
\bar{M}_2(t) = &
       + t \,{m_\pi^2 \,m_K^2\over \Delta_{\eta\pi} \,\fpid}  \,\Big(
          - \frac{2}{3} \,\bar{J}'_{\pi\eta}(0)
       \Big)
       + {m_\pi^2\over \Delta_{\eta\pi} \,\fpid}  \,\Big(
          + \frac{1}{6} \,\bar{J}_{\pi\eta}(t) \,(4 \,m_K^2 - 3 \,t)
       \Big)\nonumber\\ &
       + {1\over \Delta_{\eta\pi} \,\fpid}  \,\Big(
          + \frac{1}{4} \,\bar{J}_{\pi\pi}(t) \,(2 \,m_\pi^2 - t) \,(4 \,m_K^2 - 3 \,t)
          - \frac{1}{8} \,\bar{J}_{KK}(t) \,(4 \,m_K^2 - 3 \,t)^2
       \Big)
\nonumber\\ &
+\Delta\bar{M}_2(t)
\end{align}
with the electromagnetic contribution
\be
\Delta\bar{M}_2(t)=-\frac{1}{\epsilon_L}\,
\frac{e^2\,C }{4\sqrt3\,\fpis}\Big((4\,\mkd - 3\,t)\,\bar{J}_{KK}(t)
-4\mkd\,t\, \bar{J}'_{KK}(0)
\Big)\ .
\en
The coupling $C$ can be simply determined
from the $\pi^+-\pi^0$ mass difference,
\be
m^2_\pip-m^2_\piz = \frac{2 e^2 C}{\fpid} + O(e^2 p^2)\ .
\en
The couplings $K_i^r$ are expected to have an order of magnitude
$K_i^r (\mu=m_\rho)\sim 1/(16\pi^2)$ but, otherwise, they are not
precisely known.  Fortunately, in the $\eta\to 3\pi$ amplitude, they
appear multiplied by $e^2\mpid$. This is in contrast to other isospin
violating observables like $m^2_\Kp-m^2_\Kz$ in which they appear
multiplied by the larger factor $e^2 m_K^2$. A simple resonance saturation
estimate~\cite{Moussallam:1997xx,Ananthanarayan:2004qk} gives:
$-2K_3+K_4\simeq -4.0\cdot 10^{-3}$,  
$K_5+K_6\simeq 14.4\cdot 10^{-3}$,
$K_9+K_{10}\simeq7.5\cdot 10^{-3}$ (with $\mu=\mu_0=0.77$ GeV) which suggests
that there are cancellations in the combination relevant for $\eta\to 3\pi$
\be
-\frac{3}{2}\,(-2K^r_3+K^r_4)-K^r_5-K^r_6 +K^r_9+K^r_{10} 
+\frac{3\,C}{2\fpiq}\frac{1}{16\pi^2}(1+L_K)\simeq 0.12\cdot 10^{-3}\ .
\en

\section{Matching equations}\label{sec:appendixB} 
We reproduce below the set of matching relations from which we determine the
set of 16 polynomial parameters of the Khuri-Treiman coupled-channel equations
(i.e. eqs.~\rf{M0disp},~\rf{M2disp} and the second one of
eqs.~\rf{dispomnes}). In order to simplify the relations it is assumed here
that the chiral expressions for the $\eta\to 3\pi$ functions $\bar{M}_I$
satisfy, as in appendix~\ref{sec:appendixA}, the $w=0$ relations
$\bar{M}_1(0)=\bar{M}_2(0)=\bar{M}'_2(0)=0$. Derivatives at $w=0$ are denoted
either by dots or by primes and matrix elements of the $I=0,1$ MO matrices are
denoted here by $\Omega^{(I)}_{ij}$. The chiral LO expressions for the
$K\Kbar$ amplitudes $\bar{N}_0$, $\bar{G}_{10}$, $\bar{H}_{10}$ and
$\bar{G}_{12}$ are given in eqs.~\rf{amplitKKp2}.  A first set of five
relations is
\begin{align}
\alpha_0^N= & \bar{N}_0(0)     \nonumber\\
\alpha_0^G= & \bar{G}_{10}(0)   \nonumber\\
\alpha_0^H= & \bar{H}_{10} (0)  \nonumber\\
\alpha_2^K= & \bar{G}_{12} (0)\nonumber\\
\beta_2^K= &  \bar{G}'_{12}(0)
- (\dot{\Omega}^{(1)}_{22}+\dot{\Omega}^{(2)})\,\alpha_2^K 
\end{align}
Then, introducing the notation
\be
\bar{M}_2^{eff}= \frac{1}{2}\,\bar{M}''_2(0)-\hat{I}_2(0) 
+\left(-\dot{\Omega}^{(2)}\,\dot{\Omega}^{(1)}_{12}
-\frac{1}{2}\,\ddot{\Omega}^{(1)}_{12}\right)\,\alpha_2^K
-\dot{\Omega}^{(1)}_{12}\,\beta_2^K
\en
we have
\begin{align}
\alpha_0= & \bar{M}_0(0) +9\,s_0^2\,\bar{M}_2^{eff} 
-3\,s_0\dot{\Omega}^{(1)}_{12}\,\alpha_2^K\nonumber\\
\beta_0=  & \bar{M}'_0(0)-9\,s_0\,\bar{M}_2^{eff}
-(\dot{\Omega}^{(0)}_{11}+\dot{\Omega}^{(1)}_{11})\,\alpha_0
-\dot{\Omega}^{(1)}_{12}\,\alpha_0^G
-\dot{\Omega}^{(0)}_{12}\,\alpha_0^N
+\frac{5}{3}\,\dot{\Omega}^{(1)}_{12}\,\alpha_2^K \nonumber \\
\beta_1= & \bar{M}'_1(0)-\hat{I}_1(0)+\bar{M}_2^{eff}
\end{align}
and
\begin{align}
  \beta_0^N= & \bar{N}'_0(0)+(-\dot{\Omega}^{(0)}_{22} -\dot{\Omega}^{(1)}_{11}
  )\,\alpha_0^N-\dot{\Omega}^{(0)}_{21} \,\alpha_0-
\dot{\Omega}^{(1)}_{12} \, \alpha_0^H
\nonumber\\            
  \beta_0^G=  & \bar{G}'_{10}(0)+ (-\dot{\Omega}^{(1)}_{22}
  -\dot{\Omega}^{(0)}_{11} )\,\alpha_0^G-\dot{\Omega}^{(1)}_{21}\,\alpha_0
  -\dot{\Omega}^{(0)}_{12} \,\alpha_0^H 
\nonumber\\            
  \beta_0^H=  & \bar{H}'_{10}(0)+ (-\dot{\Omega}^{(1)}_{22}
  -\dot{\Omega}^{(0)}_{22} )\,\alpha_0^H-\dot{\Omega}^{(1)}_{21}\,\alpha_0^N
  -\dot{\Omega}^{(0)}_{21} \,\alpha_0^G 
\nonumber\\
  \gamma_0= &
  \frac{1}{2}\,\bar{M}''_0+\frac{4}{3}\,\bar{M}_2^{eff}-\hat{I}_{11}(0)
\nonumber\\ &
+(-\dot{\Omega}^{(0)}_{11} \,\dot{\Omega}^{(1)}_{11}
  -\frac{1}{2}\,\ddot{\Omega}^{(1)}_{11}
  -\frac{1}{2}\,\ddot{\Omega}^{(0)}_{11} )\,\alpha_0
+(-\dot{\Omega}^{(1)}_{11} -\dot{\Omega}^{(0)}_{11} )\,\beta_0
\nonumber\\ &
+(-\frac{1}{2}\,\ddot{\Omega}^{(1)}_{12} -\dot{\Omega}^{(1)}_{12}
  \,\dot{\Omega}^{(0)}_{11} )\,\alpha_0^G
- \dot{\Omega}^{(1)}_{12}\,\beta_0^G
\nonumber\\ &
+(-\dot{\Omega}^{(0)}_{12} \,\dot{\Omega}^{(1)}_{11}
  -\frac{1}{2}\,\ddot{\Omega}^{(0)}_{12} )\,\alpha_0^N-\dot{\Omega}^{(0)}_{12}
  \,\beta_0^N
-\dot{\Omega}^{(1)}_{12}\,\dot{\Omega}^{(0)}_{12} \,\alpha_0^H  \ 
\end{align}
where $\hat{I}_{ij}$ denote the matrix elements of the matrix sum
$\hat{\bm{I}}_a +\hat{\bm{I}}_b$ (see
eqs.~\rf{discXa}~\rf{discXb}~\rf{Ihata,b}).  The final four relations read
\begin{align}
  \gamma_0^N=  &-\hat{I}_{21}(0)+ (-\dot{\Omega}^{(1)}_{11} \,\dot{\Omega}^{(0)}_{21}
  -\frac{1}{2}\,\ddot{\Omega}^{(0)}_{21} )\,\alpha_0-\dot{\Omega}^{(0)}_{21}
  \,\beta_0- 
\dot{\Omega}^{(1)}_{12} \,\dot{\Omega}^{(0)}_{21} \,\alpha_0^G
\nonumber\\ &
+(-\dot{\Omega}^{(1)}_{11} \,\dot{\Omega}^{(0)}_{22}
-\frac{1}{2}\,\ddot{\Omega}^{(0)}_{22} -\frac{1}{2}\,\ddot{\Omega}^{(1)} 
_{11} )\,\alpha_0^N +(-\dot{\Omega}^{(0)}_{22} -\dot{\Omega}^{(1)}_{11}
)\,\beta_0^N
\nonumber\\ &
+(-\frac{1}{2}\,\ddot{\Omega}^{(1)}_{12} -\dot{\Omega}^{(1)}_{12}\, 
\dot{\Omega}^{(0)}_{22} )\,\alpha_0^H-\dot{\Omega}^{(1)}_{12} \,\beta_0^H
\nonumber\\            
 \gamma_0^G= & -\hat{I}_{12}(0)+ (-\dot{\Omega}^{(1)}_{21} \,\dot{\Omega}^{(0)}_{11}
 -\frac{1}{2}\,\ddot{\Omega}^{(1)}_{21} )\,\alpha_0-\dot{\Omega}^{(1)}_{21}
 \,\beta_0
\nonumber\\ &
+(-\dot{\Omega}^{(1)}_{22} \,\dot{\Omega}^{(0)}_{11}
-\frac{1}{2}\,\ddot{\Omega}^{(0)}_{11} -\frac{1}{2}\,\ddot{\Omega}^{(1)}_{22}
)\,\alpha_0^G+(-\dot{\Omega}^{(1)}_{22} - 
\dot{\Omega}^{(0)}_{11} )\,\beta_0^G
\nonumber\\ &
-\dot{\Omega}^{(1)}_{21}\,\dot{\Omega}^{(0)}_{12} \,\alpha_0^N
+(-\dot{\Omega}^{(1)}_{22}\,\dot{\Omega}^{(0)}_{12} -\frac{1}{2}\, 
\ddot{\Omega}^{(0)}_{12} )\,\alpha_0^H-\dot{\Omega}^{(0)}_{12} \,\beta_0^H
\nonumber\\             
 \gamma_0^H=  &-\hat{I}_{22}(0) -\dot{\Omega}^{(1)}_{21} \,\dot{\Omega}^{(0)}_{21}
  \,\alpha_0+(-\dot{\Omega}^{(1)}_{22} \,\dot{\Omega}^{(0)}_{21}
  -\frac{1}{2}\,\ddot{\Omega}^{(0)}_{21})\,\alpha_0^G-\dot{\Omega}^{(0)}_{21}
  \,\beta_0^G 
\nonumber\\ &
+(-\dot{\Omega}^{(1)}_{21} \,\dot{\Omega}^{(0)}_{22}
-\frac{1}{2}\,\ddot{\Omega}^{(1)}_{21} )\,\alpha_0^N- 
\dot{\Omega}^{(1)}_{21} \,\beta_0^N
\nonumber\\ &
+(-\frac{1}{2}\,\ddot{\Omega}^{(1)}_{22}
-\dot{\Omega}^{(1)}_{22} \,\dot{\Omega}^{(0)}_{22}
-\frac{1}{2}\,\ddot{\Omega}^{(0)}_{22} )\, 
\alpha_0^H+(-\dot{\Omega}^{(1)}_{22} -\dot{\Omega}^{(0)}_{22} )\,\beta_0^H
\nonumber\\
  \gamma_2^K=  &-\hat{I}_2^K(0)+ (
  -\frac{1}{2}\,\ddot{\Omega}^{(1)}_{22} -\dot{\Omega}^{(2)}\,
\dot{\Omega}^{(1)}_{22} -\frac{1}{2}\,\ddot{\Omega}^{(2)})\, 
\alpha_2^K\nonumber\\
& +(-\dot{\Omega}^{(1)}_{22} -\dot{\Omega}^{(2)})\,\beta_2^K\ .            
\end{align}

\section{Angular integrals and hat functions}\label{sec:appendixC} 
Using the dispersive representations~\rf{disprelMI} of the functions
$M_I$, one can express the
angular integrals in the following form which  displays explicitly their
singularity when $s \to (\meta-\mpi)^2$. For $I=0,2$ one has
\bea 
&&\braque{M_I}(s)=\tilde{\alpha}_I +{1\over2}\tilde{\beta}_I(3s_0-s)
+{2R_I^0(s)\over\kappa(s)} 
-{1\over\pi}\int_{4\mpid}^\infty dt'\, \disc[M_I(t')]\,K^{(0)}(t',s)\nonumber\\
&&\braque{zM_I}(s)={1\over6}\tilde{\beta}_I\kappa(s) 
+{4R_I^1(s)\over(\kappa(s))^2 } 
-{1\over\pi}\int_{4\mpid}^\infty dt'\, \disc[M_I(t')]\,K^{(1)}(t',s)
\ena
where $\kappa(s)$ is given in eq.~\rf{tuandkappa}. For $I=1$ one has
\be
\braque{z^nM_1}(s)=
\left({2\over\kappa(s)}\right)^{n+1} R_1^n(s)
-{1\over\pi}\int_{4\mpid}^\infty dt'\,\disc[M_1(t')]\,P^{(n)}(t',s)
\en
The functions $R_I^n(s)$ which control the singularities arise from the part
of the $t$ integration contour which encircle the unitarity cut, they are
given by 
\be
R_I^n(s)=i\int_{4\mpid}^{\re[t^-(s)]} dt'\, (t'-\frac{1}{2}(3s_0-s))^n\,
\disc[M_I(t')]
\en 
(where $t^\pm(s)$ is given in eq.~\rf{t+-}) in the $s$ range
\be
\frac{1}{2}(\metad-\mpid) < s < \metad-5\mpid
\en
and $R_I^n(s)=0$ otherwise. In particular, no divergence occurs when $s\to
4\mpid$ or $s\to (\meta+\mpi)^2$.

The kernels which are needed here are given by the following expressions
\bea
&& P^{(0)}(t',s)={1\over t'} + {1\over\kappa(s)} L(t',s)\nonumber\\
&& P^{(1)}(t',s)={2\over\kappa(s)}+{(2t'+s-3s_0)\over (\kappa(s))^2}\, L(t',s)
\nonumber\\
&& P^{(2)}(t',s)={1\over3 t'}+{2(2t'+s-3s_0)\over\kappa(s)^2}
+{(2t'+s-3s_0)^2\over\kappa(s)^3}\,L(t',s)\ .
\ena
and
\bea
&& K^{(0)}(t',s)= {3s_0-s\over 2 (t')^2} + P^{(0)}(t',s)\nonumber\\
&& K^{(1)}(t',s)= {\kappa(s)\over 6(t')^2} +  P^{(1)}(t',s)\ .
\ena
The function $L(t',s)$ arises from the parts of the $t$ integration contour not
taken into account in the functions $R_I^n$ (the function $L(t',s)$ thus vanishes when
$s\to (\meta-\mpi)^2$, such that the kernels remain finite) it is given by
\begin{itemize}
\item[1)] $4\mpid\le  s < (\meta-\mpi)^2$: 
\be
L(t',s)=\log(t^+(s)-t'+i\epsilon)-\log(t^-(s)-t'+i\epsilon)
\en
\item[2)] $(\meta-\mpi)^2\le s< \metad-5\mpid$:
\bea
&&L(t',s)=\log(t^+(s)-t')-\log(\undemi(3s_0-s)-t'+i\epsilon)\nonumber\\
&&-\log(t^-(s)-t')+\log(\undemi(3s_0-s)-t'-i\epsilon)
\ena
\item[3)] $\metad-5\mpid\le s < \infty$
\be
L(t',s)=\log(t'-t^+(s))-\log(t'-t^-(s))\ .
\en
\end{itemize}

\section{Supplementary material}\label{sec:supplement}
The integral equations for the amplitudes $M_I$ depend linearly on the 16
polynomial parameters. The matching equations are also linear.  One can then
express the amplitudes in which, for instance, the four leading parameters
$\alpha_0$, $\beta_0$, $\gamma_0$, $\beta_1$ are fixed to some arbitrary
values and the remaining 12 parameters are fixed from the corresponding
matching equations in the form of a linear superposition
\begin{align}\lbl{polylin}
M_I(s) = & M_I^{(0000)}(s) 
+\alpha_0 M_I^{(1000)}(s) 
+\beta_0  M_I^{(0100)}(s)\nonumber \\ &
+\gamma_0 M_I^{(0010)}(s)
+\beta_1  M_I^{(0001)}(s)\ .
\end{align} 
We provide our numerical results for the amplitudes in which the parameters
$\alpha_0$, $\beta_0$, $\gamma_0$, $\beta_1$ are either 0 or 1 in five data
files: MI\_0000.dat, MI\_1000.dat, MI\_0100.dat, MI\_0010.dat,
MI\_0001.dat. In each file, the first column is $s$ (in $\hbox{GeV}^2$) and
the other columns are the corresponding real and imaginary parts of $M_0$,
$M_1$ and $M_2$. We note that these amplitudes depend on the energy value
$E_{asy}$ above which the $T$ matrix parameters are set to their asymptotic
values. We take here $E_{asy}=10$ GeV. We give below the corresponding values
of the four polynomial parameters for several cases considered in this
paper.
\begin{itemize}
\item[1)] Chirally matched amplitude, $L_3=-3.82\cdot 10^{-3}$:\\[-0.2cm] 
\be\ba{lrl}
 \alpha_0=&   -0.69285 & +i\,0.05692\\
 \beta_0 =&   17.27894 & -i\,0.64122    \\
 \gamma_0=&  -46.42237 & +i\,1.02473    \\
 \beta_1 =&    8.45260 & +i\,0.27853   \\ 
\ea
\en 

\item[2)] Chirally matched amplitude, $L_3=-2.65\cdot 10^{-3}$:\\[-0.2cm] 
\be\ba{lrl}
 \alpha_0=&  -0.67534   & +i\, 0.05677  \\
 \beta_0 =&   17.0817 &   -i\, 0.63953  \\
 \gamma_0=&  -42.5778 &   +i\, 1.02438  \\
 \beta_1 =&    6.7383 &   +i\,  0.27776 \\ 
\ea
\en 

\item[3)] Fitted amplitude:
\be\ba{lrl}
 \alpha_0=&  -0.67534    +i\, 0.05677  \\
 \beta_0 =&   17.2280    -i\, 0.63953  \\
 \gamma_0=&  -44.3672    +i\, 1.02438  \\
 \beta_1 =&   6.53640    +i\, 0.27776  \\ 
\ea
\en 
Note that $ \alpha_0$ and the imaginary parts of $\beta_0$, $\gamma_0$,
$\beta_1$ are the same as in 2).
The covariance matrix of the three fitted parameters reads

\def\arraystretch{0.60}%
\bt{c|ccc}\\ 
\BB \         & $\beta_0$  & $\gamma_0$  & $\beta_1$ \\ \hline
\TT $\beta_0$ &  0.261     &             &          \\
\TT $\gamma_0$& -0.846     &  2.847      &          \\
\TT $\beta_1$ &  0.439     & -1.439      & 0.742    \\
\et
\vspace{0.2cm}

and, finally, the value of the normalisation parameter $\lambda$ (see
eq.~\rf{chi2def}) is: $\lambda=1.009\pm0.001$.
\end{itemize}

\end{document}